\documentclass[acmtog,nonacm]{acmart}
\setcopyright{none}
\renewcommand\footnotetextcopyrightpermission[1]{}
\usepackage{booktabs} %

\graphicspath{{"../"}}

\usepackage[ruled]{algorithm2e} %

\SetAlFnt{\small}
\SetAlCapFnt{\small}
\SetAlCapNameFnt{\small}
\SetAlCapHSkip{0pt}

\acmJournal{TOG}

\newcommand{\etal}{\textit{et al.}}

\newcommand{\point}{p} %

\newcommand{\loss}{\mathcal{L}} %

\DeclareMathOperator{\bce}{BCE} %

\usepackage[labelsep=period,skip=1pc,size=small]{caption}
\usepackage{subcaption}
\usepackage{booktabs}
\usepackage{wrapfig}
\usepackage{enumitem}

\usepackage{soul}

\newcommand{\Tab}[1]{Table~\ref{tab:#1}}

\definecolor{turquoise}{cmyk}{0.65,0,0.1,0.3}
\definecolor{purple}{rgb}{0.65,0,0.65}
\definecolor{dark_green}{rgb}{0, 0.5, 0}
\definecolor{orange}{rgb}{0.9, 0.6, 0.1}
\definecolor{red}{rgb}{0.8, 0.2, 0.2}
\definecolor{darkred}{rgb}{0.6, 0.1, 0.05}
\definecolor{blueish}{rgb}{0.0, 0.3, .6}
\definecolor{light_gray}{rgb}{0.7, 0.7, .7}
\definecolor{pink}{rgb}{1, 0, 1}
\definecolor{greyblue}{rgb}{0.25, 0.25, 1}
\definecolor{amethyst}{rgb}{0.6, 0.4, 0.8}

\newcommand{\Lone}{{L_1}}

\newcommand{\mylesdata}{Myles \etal~\shortcite{Myles16}\xspace}

\newcommand{\customparskip}{0.5em}
\renewcommand{\paragraph}[1]{\vspace{\customparskip}\noindent\textbf{#1}}

\usepackage{multicol}

\usepackage{longtable}  %

\usepackage{lipsum}

\newenvironment{parWithWrapFigure} %
{\begingroup
\setlength{\columnsep}{1em}%
\setlength{\intextsep}{0em}%
\setlength{\arraycolsep}{0pt}} %
{

\endgroup}

\begin{document}

\title{NESI: Shape Representation via Neural  Explicit Surface Intersection }

\author{Congyi Zhang}
\affiliation{%
\institution{University of British Columbia}
\country{Canada}}
\author{Jinfan Yang}
\affiliation{
\institution{University of British Columbia}
\country{~}}
\author{Eric Hedlin}
\affiliation{
\institution{University of British Columbia}
\country{~}}

\author{Suzuran Takikawa}
\affiliation{%
  \institution{University of British Columbia}
  \country{~}}
\author{Nicholas Vining}
\affiliation{%
  \institution{NVIDIA}
  \country{~}}
\author{Kwang Moo Yi}
\affiliation{%
\institution{University of British Columbia}
\country{~}}
\author{Wenping Wang}
\affiliation{%
\institution{University of Texas A\&M}
\country{~}}  
\author{Alla Sheffer}
\affiliation{%
\institution{University of British Columbia}
\country{~}}  

\begin{teaserfigure}
\begin{center}
	\includegraphics[width=\linewidth]{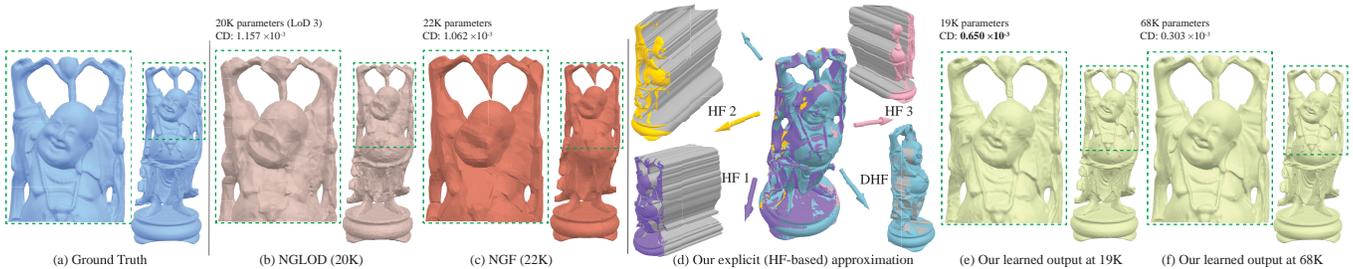}
\end{center}
\caption{NESI compactly represents detailed 3D shapes (a, 109K faces) as intersections of neural explict height-field surfaces (HFs)  (d-f). It compactly encodes a set of differently oriented double (DHF) and single HFs (d) whose intersection well approximates the input. Given the same parameter budget, NESI provides more accurate approximations (e) of the inputs than state of the art alternatives: (b)  NGLOD \cite{nglod}, (c) NGF \cite{ngf}. At higher parameter counts (f) our results are visually indisitinguishable from the input.}
\label{fig:teaser}
\end{teaserfigure}

\begin{abstract}
Compressed representations of 3D shapes that are compact, accurate, and can be processed efficiently directly in compressed form, are extremely useful for digital media applications.
Recent approaches in this space focus on learned {\em implicit} or {\em parametric} representations. While implicits are well suited for tasks such as in-out queries, they lack natural 2D parameterization, complicating tasks such as texture or normal mapping. Conversely, parametric representations support the latter tasks but are ill-suited for occupancy queries. 
We propose a novel learned alternative to these approaches, based on intersections of localized {\em explicit}, or {\em height-field}, surfaces. Since explicits can be trivially expressed both implicitly and parametrically, NESI directly supports a wider range of processing operations than implicit alternatives, including occupancy queries and parametric access. 
We represent input shapes using a collection of differently oriented height-field bounded half-spaces combined using volumetric Boolean intersections. We first tightly bound each input using a pair of oppositely oriented height-fields, forming a {\em Double Height-Field (DHF) Hull}. We refine this hull by intersecting it with additional {\em localized height-fields (HFs)} that capture surface regions in its interior. 
We minimize the number of HFs necessary to accurately capture each input
and compactly encode both the DHF hull and the local HFs as neural functions defined over subdomains of $\mathbb{R}^2$. This reduced dimensionality encoding delivers high-quality compact approximations. 
 Given similar parameter count, or storage capacity, NESI significantly reduces approximation error compared to the state of the art, especially at lower parameter counts. 
\end{abstract}

\maketitle
\section{Introduction}
\label{sec:intro}
Shape representations which support efficient geometry manipulation and processing, while also being accurate and compact, are of major interest for applications such as video games, 3D content streaming, and VR/AR \cite{Karis:NaniteTalk}.
Popular traditional representations include implicit surfaces, piecewise explicit representations, and piecewise parametric surfaces (B-Reps) \cite{farin,Botsch,DannyBook}, each with pros and cons. Implicits support in-out queries but cannot easily be parameterized, and thus do not directly support important geometry processing tasks such as texture mapping.  
Piecewise parametric or piecewise explicit surface representations, including meshes, can be effectively used for many geometry processing operations \cite{Botsch,DannyBook}, but do not directly support in-out queries. 
In general, traditional representations are far from compact, and require large numbers of parameters, or degrees of freedom, to capture detailed shapes; this has motivated the recent quest for more compact neural alternatives (Sec. ~\ref{sec:related}). State-of-the-art neural implicit  ~\cite{nglod,vqad,siren,takikawa2023compact} or parametric \cite{ngf,Morreale2022NCS,Morreale2021NSM} shape representations provide a compact alternative to traditional representations, and can accurately encode highly detailed shapes using much fewer parameters. However, they inherit the processing limitations of their traditional counterparts: neural implicits do not support operations that require local or global surface parameterization, such as meshing and texture mapping, while neural parametric surfaces do not support occupancy queries. 
We propose a novel  shape representation which is more compact than existing alternatives, and supports both fast in-out queries and processing tasks that leverage parameter domain information, such as texture or normal mapping. 

\begin{parWithWrapFigure}
We achieve this goal by leveraging the representational power of explicit, or {\em height-field} (HF), surfaces.
We recall that an explicit or height-field (HF) surface is defined as the graph of a function $z=f(x,y)$ over a 2D domain $\Omega \in \mathbb{R}^2$ (see inset) and has an explicit parameterization relative to this domain (i.e. $P(x,y)= (x, y, f(x,y))$. Moreover, HF surfaces partition space into inside (light blue) and outside (white) {\em half-spaces}: a point $(x,y,z) \in \mathbb{R}^3$ is inside the half-space, or volume, $E(f)$, associated with the HF surface $f$ if and only if $(x,y)\in \Omega$ and $f(x,y) > z$.
\begin{wrapfigure}{r}{.2\columnwidth}
    \includegraphics[width=\linewidth]{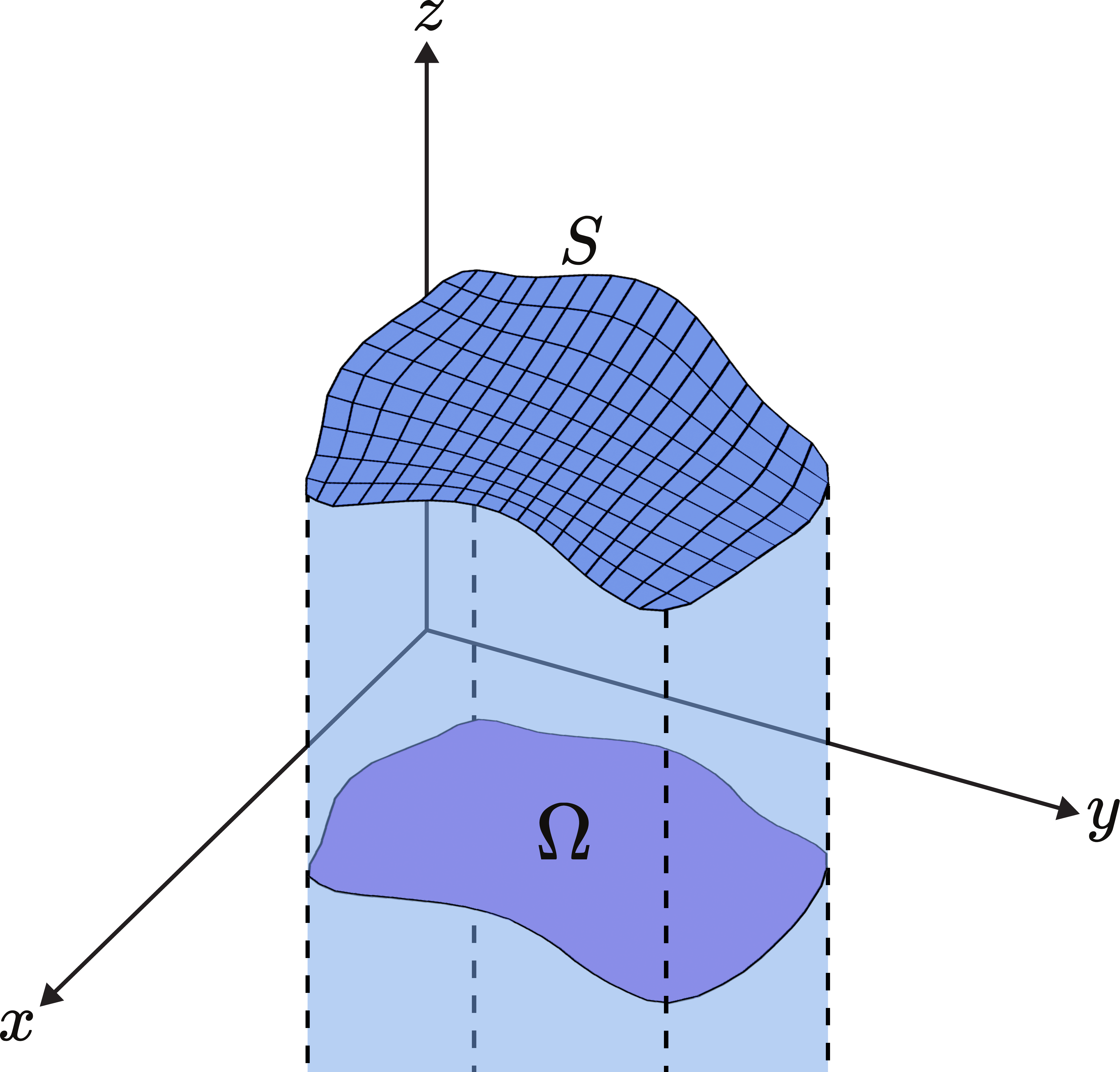}
\end{wrapfigure}
As such, explicit HF-based representations combine the processing advantages of implicit and parametric ones. They are also inherently more compact than general parametric or implicit representations due to dimensionality reduction: all one must store are the $z=f(x,y)$ values over their parameter domain $\Omega$. However, the range of shapes representable by a single HF surface is highly limited, as an HF can only represent a surface with a single $z$ value for each $(x,y)$. Piecewise explicit surface \cite{guskov2000normal,maggiordomo2023micro} or volume \cite{yang2020dhfslicer,Muntoni2019,Muntoni2018} representations, which approximate surfaces or volumes using unions of HF surface patches or bounded half-spaces, are no longer compact and cannot robustly support in-out queries (Sec. ~\ref{sec:related}). 
\end{parWithWrapFigure}

\begin{figure}
\centering
\includegraphics[width=.92\linewidth]{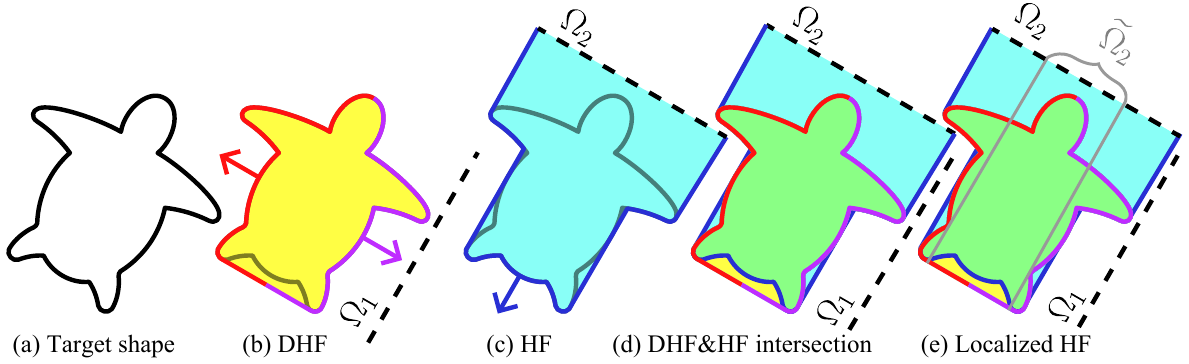}
\caption{A shape (a) represented (d) as the intersection (green) of a DHF hull (b) and an additional HF (c). Localizing the HF to a narrower parameter domain $\Omega$ (e), by implicitly assuming it to match the DHF elsewhere, reduces representation redundancy.} 
\vspace{-5pt}
\label{fig:nesi}
\end{figure}

We extend the advantages of explicit representations to generic shapes by observing that even extremely complex shapes can be accurately approximated using a Boolean {\bf intersection} of just a few judiciously selected overlapping HF half-spaces; for instance, the {\em happy buddha} (Fig. ~\ref{fig:teaser}d) can be accurately approximated by intersecting just five such half-spaces. Moreover, using intersecting half-spaces as a shape representation allows for robust and efficient in-out queries {\em and} trivial surface parameterization. We refer to this HF intersection based representation as {\em Explicit Surface Intersection}, or {\em ESI}.
We further note, importantly, that an intersection of HF half-spaces can be compactly encoded in neural form by taking advantage of the fact that each HF is simply defined by a function $z=f(x,y)$ over a 2D domain; encoding HFs in this manner produces a {\em Neural} Explicit Surface Intersection, or {\em NESI}, representations.  

While early attempts at representing shapes using HF intersections \cite{shade1998layered,Richter2018Matryoshka} use a large set of fixed, shape-independent, HF half-space orientations, or axis directions, they frequently fail to approximate large portions of the input surfaces 
(see Sec.~\ref{sec:related}, Fig.~\ref{fig:fixed}).
In contrast, we compute a minimal set of best-approximating HF axes per input, achieving high approximation quality with just a handful of HFs (Sec~\ref{sec:construction}, Fig.~\ref{fig:fixed}). 
This ability to accurately represent diverse geometries using a small handful of HFs is key to our shape representation.

Given an input shape defined via a mesh or other standard representation, we tightly bound it using a pair of oppositely oriented HF half-spaces that jointly define a {\em Double HF (DHF) hull} of the input shape (Fig.~\ref{fig:nesi}b, Fig.~\ref{fig:teaser}d - blue).   
We refine this hull by intersecting it with additional HF half-spaces that capture input surface regions lying inside the hull's interior (Fig.~\ref{fig:nesi}c, Fig.~\ref{fig:teaser}d - yellow, purple, pink). We optimize the choice of DHF and HF axis directions to minimize approximation error, while still keeping the number of HFs used as small as possible. Performing this optimization via brute-force search makes the problem intractable, as even the evaluation of the approximation quality of a single ESI is highly time consuming. We make the problem tractable via a combination of pre-computation and a branch-and-bound discrete optimization strategy that quickly rejects direction candidates to arrive at an optimal solution with minimal HF count.
We avoid representational redundancy, where multiple explicits describe the same areas on the input shape (e.g. turtle arms in Fig.~\ref{fig:nesi}c), by only storing HF surface geometry in areas where it is not already adequately described by other explicits (Fig.~\ref{fig:nesi}e). This 
{\em localization} process reduces the geometric complexity of each HF, and thus the number of parameters required to encode it. 
Using our scheme, the vast majority of shapes in commonly used 3D shape databases \cite{abcdataset,thingi10k} can be accurately represented using a DHF hull plus one to three additional HFs, with many shapes accurately represented using their DHF hull alone (31\% of objects tested in our experiments; Sec.~\ref{sec:results}). 
We encode our DHFs and HFs using SIREN Multi-Layer Perceptron (MLP) architecture~\cite{siren}, as it does not require positional encoding~\cite{tancik2020fourier} and yet is able to encode both high- and low-frequency shape details.
Our experiments (Sec. ~\ref{sec:use},~\ref{sec:results}) demonstrate that NESI allows for straightforward surface parameterization, enabling texturing (Fig. ~\ref{fig:applications}) and other similar tasks, and supports efficient and accurate in-out queries performed by following the sequence of local intersection operations; the latter are used to ray-trace all of our outputs throughout the paper.

We thoroughly validate the effectiveness of our method by evaluating ESI accuracy across 320 inputs, and by learning NESI representations of 
100 diverse representative shapes using four different parameter counts for each shape.
We compare our results to those generated by leading alternatives using same or higher parameter counts. 
On average our outputs are 30\% more accurate than those produced by the best-performing alternative (NGF \cite{ngf}) using the same or lower parameter counts, with improvement most pronounced at lower parameter counts. 86\% of our learned outputs more accurately approximate the input ground truth shapes than those produced by this alternative, and our largest error across all inputs tested is only one third of theirs.

\begin{figure}
\includegraphics[width=\linewidth]{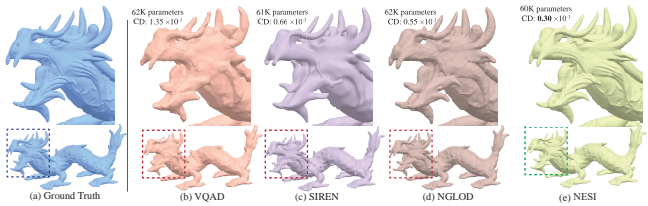}
\caption{NESI approximations (e) of complex inputs (a) are much more detailed and accurate than those generated by leading implicit alternatives: (b) \cite{vqad}, (c) SIREN \cite{siren}, and (d) \cite{nglod},  despite using fewer parameters.}
\label{fig:implicits}
\end{figure}

\section{Background and Related Work}
\label{sec:related}
A vast body of previous work exists on 3D shape representations, each with their pros and cons. Here, we focus on representations closest to NESI in terms of goals or properties.

\paragraph{Traditional Shape Representations.} 
{\em Parametric}, or boundary (B-Rep) representations, including polygonal meshes, define the bounding surfaces of closed 3D shapes using collections of parametric patches connected together along common boundary seams \cite{farin,Botsch}. While well suited for surface-based tasks such as texturing, meshing, or (re)parameterization, computing in-out queries using these representations requires costly intersection computations and auxiliary data structures, making them less suitable for tasks such as raytracing or collision detection. 
Accurately approximating input shapes using either meshes or piecewise smooth parametric patches  requires large patch and parameter counts \cite{luebke2001developer,Botsch,LevyT,Litke}. 
Mesh compression schemes target compact mesh storage and transmission, and require decompressing the outputs prior to actual use \cite{alliez2005recent,maglo20153d}.
Geometry Images ~\cite{geometryimages,sander2003multi,Carr:2006:RMG} compress meshes via 2D parameterization; they 
retain the inherent limitations of parametric representations and require large amounts of atlas space to obtain quality approximations. Our NESI representation combines the advantages of parametric representations with fast in-out queries, has a much smaller memory footprint, and can be processed directly in its compressed form. %
In Fig \ref{fig:overview} we approximate the Buddha mesh with ~120K triangles using just 48k parameters, producing a visually practically identical render (the chamfer distance between our model and the input mesh is 0.28).

{\em Implicit} representations (e.g. ~\cite{BlinnImplicits, osher2004level,wyvill1998blob}) define a closed surface $S$ as a level set of a function $F:\mathbb{R}^3 \to \mathbb{R}$.
Implicits support efficient in-out queries~\cite{jones20063d,Takikawa2022SDF} but are difficult to parameterize
either globally or locally \cite{schmidt2006interactive}, making them challenging to texture, mesh, or normal map.
Converting generic surfaces into analytic implicit form remains an open problem \cite{reverseengsurvey}; the commonly-used grid-based representations of implicits (e.g. \cite{musethvdb}) are highly memory consuming.

\paragraph{Explicit Surfaces.}
Classical {\em explicit}, or height-field (HF), surfaces are defined as height functions $z = F(x,y), ~(x,y) \in \Omega$ over a 2D domain $\Omega \in \mathbb{R}^2$ \cite{farin}. In the general case, the parameter domain can lie in any plane in $R^3$, and the height $z$ represents the offset or distance from this plane along the plane's normal, or {\em axis}. HFs can be viewed as a special case of parametric surfaces and trivially support parameterization-based tasks. Since  few shapes can be described by a single explicit surface, numerous attempts had been made to describe shapes using combinations of multiple HFs. 

\begin{parWithWrapFigure}
\begin{wrapfigure}{l}{.3\columnwidth}%
\includegraphics[width=\linewidth]{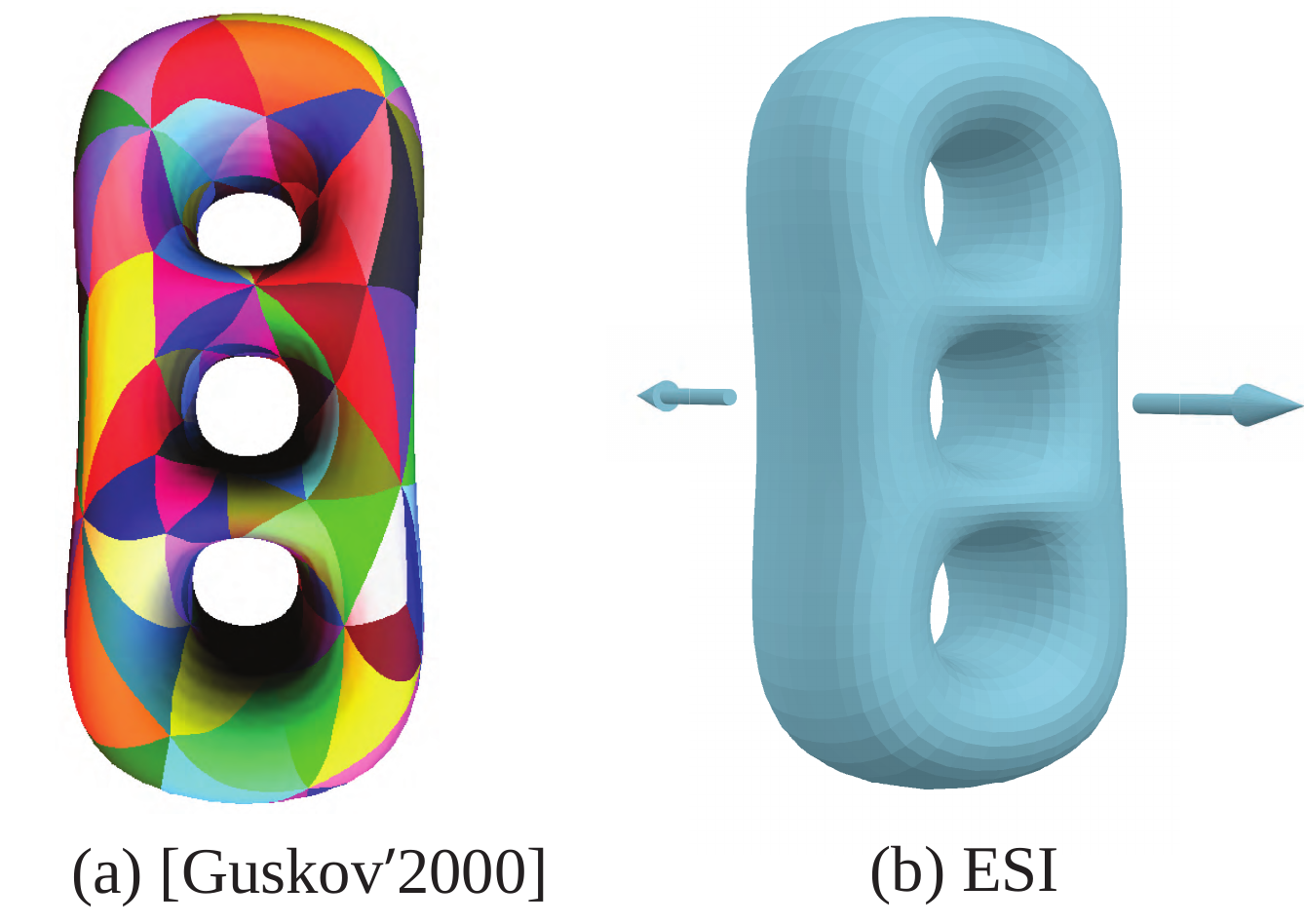}%
\end{wrapfigure}
Approximating existing shapes using piecewise HFs defined over polygonal domains \cite{khodakovsky2000progressive,guskov2000normal,novak2012rasterized} significantly reduces the memory footprint of a shape relative to a standard mesh representation, and facilitates efficient rendering using displacement maps \cite{maggiordomo2023micro,thonat2021tessellation}.
Accurate piecewise explicit approximation of complex shapes requires a large number of patches and is far from compact. For instance, \cite{guskov2000normal} uses 98 patches to approximate the three-holed torus (inset, right), whereas we approximate it using a single DHF hull (inset, left); %
\cite{novak2012rasterized} use hundreds of patches to represent the dragon in Fig.~\ref{fig:implicits}, which we approximate using one DHF and 3 HFs. 
\end{parWithWrapFigure}

\begin{parWithWrapFigure}
\begin{wrapfigure}{l}{.4\columnwidth}%
\includegraphics[width=\linewidth]{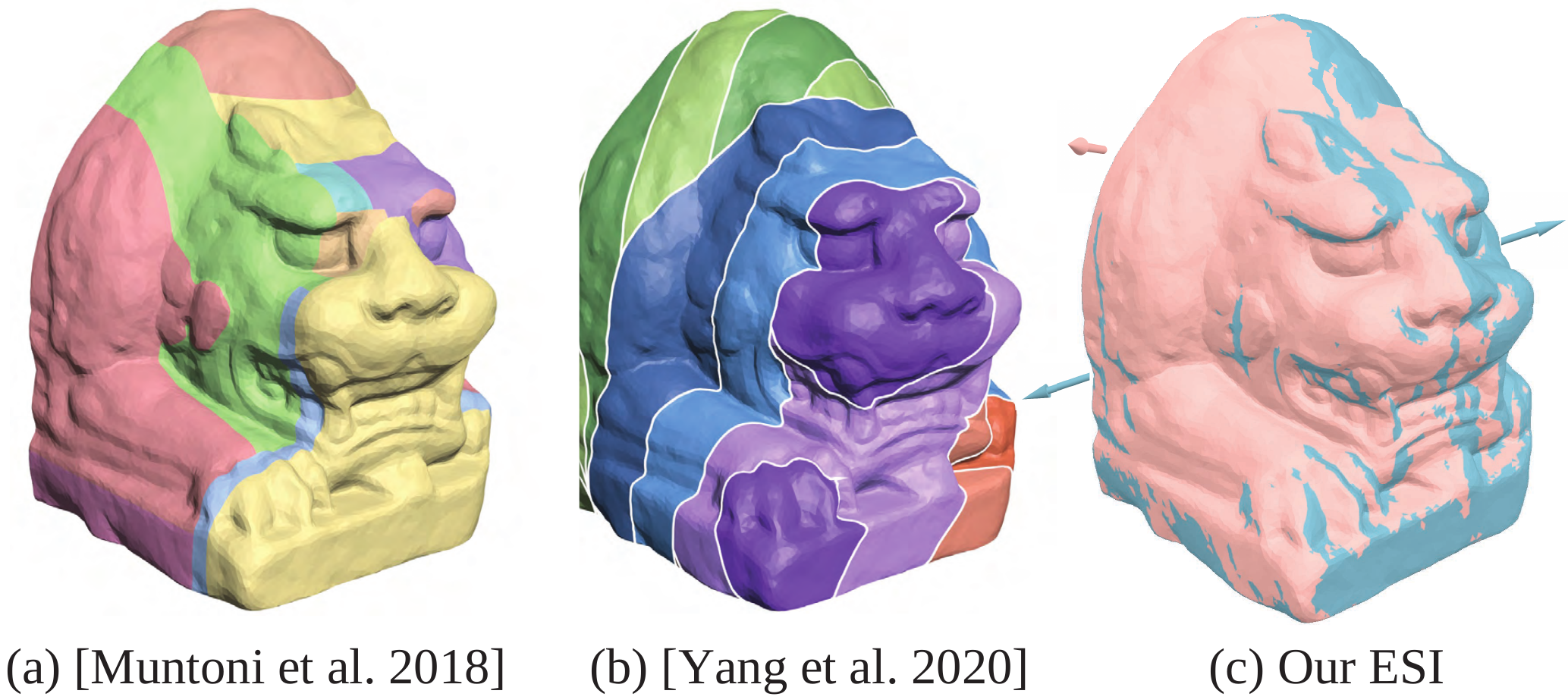}%
\end{wrapfigure}
By defining the ``inside'' of a height-field $z=f(x,y)$ as the {\em volume} between the parameter domain and the surface, explicits can also be viewed as a special case of occupancy function implicits \cite{occnet}: points $(x,y,z)$ are inside the shape if and only if $(x,y) \in \Omega$ and $z \in [0,f(x,y)]$ (placing the parameter domain at $z=-\infty$ partitions $\mathbb{R}^3$ into inside and outside half-spaces). Several fabrication methods partition shapes into explicit volumes bounded by either a height-field and its parameter domain \cite{FeketeMiller,Hu, herholz2015approximating, gao2015revomaker, Muntoni2018, Muntoni2019} (inset, a), or by pairs of oppositely oriented height-fields ~\cite{yang2020dhfslicer,alderighi2021volume} (inset, b). 
Both piecewise and volumetric explicit representations are far from compact, requiring high block counts for quality approximation; for the example  Muntoni et al. \shortcite{Muntoni2018} require 13 explicit volumes and \cite{yang2020dhfslicer} requires 12 to approximate the lion statue in the inset; we accurately approximate this input with one DHF hull and one HF (inset, c). More importantly, while partition-based representations are theoretically suitable for in-out queries, in practice assessing if a point is inside a union of non-overlapping blocks leads to false negatives for points next to boundaries between the different volumes, even deep inside the original shape. The likelihood of such catastrophic failures increases when the individual explicit volume geometries are compressed. %
By representing shapes as intersections of explicits, rather than unions, we drastically reduce the number of explicits required to accurately represent general shapes and sidestep the need to handle gaps and floating point issues along internal boundaries.
\end{parWithWrapFigure}%

\begin{figure}
\includegraphics[width=.9\linewidth]{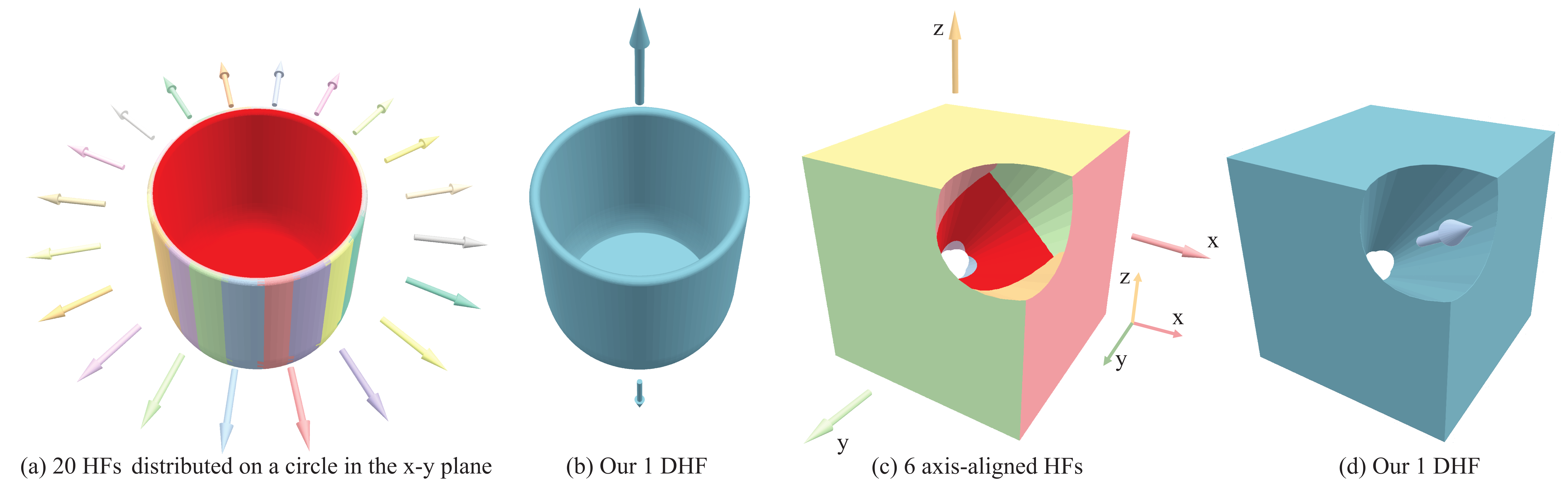}
\caption{Depth fusion methods (a,c) use intersections of  multiple depth maps with fixed, input independent axis directions (visualized by arrows) to approximate input shapes. These methods often fail to approximate large parts of the input surfaces (red in a and c) not visible along these axes. (a) 20 axis directions evenly distributed on a circle in the $x-y$ plane \cite{shade1998layered}; (c) 6 directions aligned with the +/- axes of the object's coordinate system \cite{Richter2018Matryoshka}.  (b,d) ESI accurately captures both shapes using a single DHF with automatically computed optimal axis.}
\label{fig:fixed}
\end{figure}

Our representation is inspired by depth fusion approaches for shape reconstruction \cite{turk1994zippered,Curless1996range,Richter2018Matryoshka} and representation \cite{shade1998layered,Richter2018Matryoshka} (Fig.~\ref{fig:fixed}ac). These methods define shapes as intersections of differently oriented depth maps or height-fields. The key difference between these approaches and ours is the choice of HF orientations, or axis directions.  Depth fusion methods rely on large sets of {\em input independent} depth map axis directions. Reconstruction methods such as \cite{turk1994zippered,Curless1996range} use input camera views as directions. Others rely on a fixed set of axis directions: e.g. \cite{shade1998layered} uses 20 directions evenly distributed on a circle in the $xy$ plane (Fig.~\ref{fig:fixed}a), while \cite{Richter2018Matryoshka} uses the positive and negative axes of the standard Euclidean coordinate system (Fig.~\ref{fig:fixed}c).  As \cite{Richter2018Matryoshka} acknowledges, this approach often fails to capture large portions of input shape surfaces. To address this challenge, they use 5 layers of depth maps, placed one inside the other (forming a ``matryoshka''); effectively, this means their method requires 30 (5x6) depth maps. 
The key distinction between these works and NESI is our use of HF axes that are {\em optimized per-input} so as to maximize approximation quality (Fig~\ref{fig:fixed}bd). Finding these optimal directions efficiently requires solving a complex combinatorial optimization problem across a large potential solution space (Sec~\ref{sec:construction}). This optimization based approach allows us to approximate 3D shapes with much higher accuracy, while using significantly fewer HFs overall. We require only 1 DHF for the examples in Fig~\ref{fig:fixed};  on average we use 1 DHF and fewer than 2 HFs to well approximate the ~320 inputs tested (Sec~\ref{sec:results}).

\paragraph{Neural Shape Representations.} 
Recent research efforts have attempted to encode many of the representations above using neural networks. 
Learning meshes or general explicit/parametric B-Rep/patch-based representations is known to be challenging due to their topological irregularity \cite{MeshCNN,maron} and the need to ensure continuity across inter-patch boundaries \cite{groueix2018}. 
AtlasNet \cite{groueix2018} and its followups \cite{deprelle2019learning,deprelle2022learning,deng2020,bednarik2020} represent surfaces using disconnected partially overlapping patches. Since this representation is not watertight, it cannot be reliably used for in-out queries.
Yang et al. ~\shortcite{Yang2023neuralparametric} rely on largely manually created patch layouts to learn closed B-reps of input shapes.
Both families of methods require megabytes of storage (20 for \cite{deng2020,bednarik2020} and 5 for Yang et al.) to accurately represent input shapes. We achieve higher accuracy  using filesizes under 280 kilobytes (Sec~\ref{sec:results}). 
Moreover, NESI is computed fully automatically and robustly supports in-out queries.

Neural Surface Maps \cite{morreale2021neuralsurfacemaps} represents surfaces as learned geometry images; follow-up work \cite{Morreale2022NCS} learn geometry images and a series of patch-based displacements. 
Both methods suffer from the same issues as classical geometry images, most notably lack of support for in-out queries and a requirement that the input surface be cut so that it is homeomorphic to the unit disk. 
Neural displacement methods \cite{chen2023neural,ngf} (Fig~\ref{fig:teaser}c) represent surfaces as a simplified coarse meshes overlaid with a neurally encoded displacement map; like other displacement map based representations, they do not directly support in-out queries, and require a sufficiently dense base mesh to capture topological details. Since extreme mesh simplification can be challenging for inputs with complex topology, these methods may fail or introduce severe visual artifacts at low parameter counts (Sec.~\ref{sec:results}).  We outperform the state-of-the-art methods in this category \cite{ngf,Morreale2022NCS} by notable margins (Sec.~\ref{sec:results}). NESI remains robust across all 400+ inputs and parameter count combinations tested.

Many methods address learning of compact neural {\em implicit} shape representations, including occupancy maps \cite{occnet} and Signed Distance Functions~(SDF)~\cite{instantngp,ChenSDFs}. %
Recent efforts include learning compact neural implicit functions with a variety of internal representations~\cite{deepsdf, occnet, imnet,sitzmann2019metasdf,chen2023neurbf,idf,siren,davies2021effectiveness,spheres}, or focusing on adaptive multiresolution hierarchies, combining sparse hierarchical grids with neural networks \cite{nglod, vqad}. Yifan et al. ~\shortcite{idf} represent shapes as a combination of an implicit SDF and a height, or displacement map. While drastically more efficient than naive storage, these methods inherit the limitations of traditional implicits when it comes to parameterization-driven processing tasks such as texturing or geodesic computation. Sec. ~\ref{sec:results} compares NESI to representative recent works in this category \cite{nglod, vqad,siren,idf,chen2023neurbf,spheres,ngf};  our method outperforms the best performing implicit-based alternative (SIREN w/o eikonal loss) on 93\% of the input shape and parameter count combinations tested, and improves accuracy (measured using $L_{1}$ Chamfer disneuratance) by a factor of 2.6.

\cite{Richter2018Matryoshka} propose a neural encoding of their depth map grid based shape representation. They use 30 (5x6) depth maps to encode each shape. Their reliance on grids ($256^3$ in their implementation) limits the accuracy of their outputs. NESI representation is based on precise HF/DHF intersection, and uses smooth 2D basis functions to encode the individual explicits, facilitating a much higher degree of accuracy. See Sec.~\ref{sec:results} for additional comparisons. 

Lastly, several methods focus on compact neural representation of specific classes of shapes, e.g. CAD models ~\cite{Yu2023CSG,Lin2022neuform}. NESI is not class specific, and as demonstrated in Sec.~\ref{sec:results} it compactly and accurately represents both organic and CAD shapes.

\begin{figure*}
\centering
\includegraphics[width=.9\textwidth]{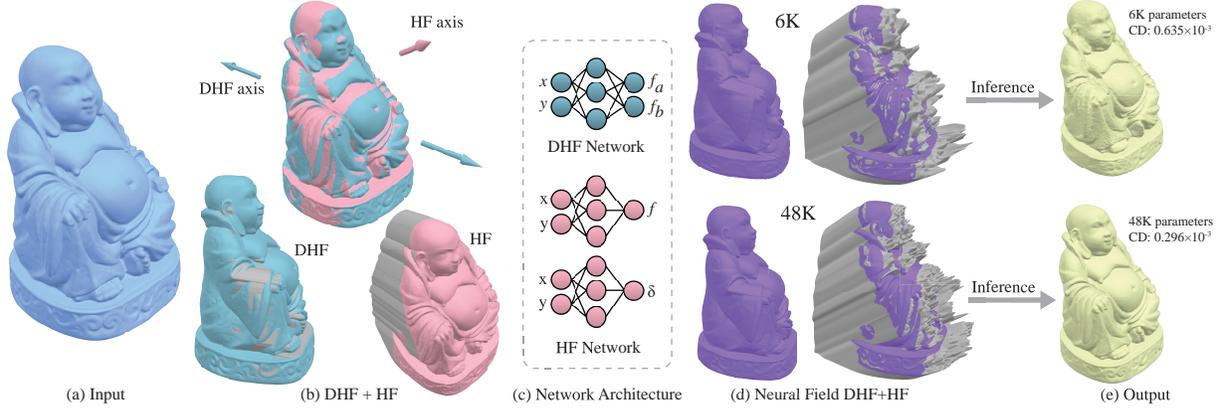}
\caption{Given an input 3D shape (a), we compute a DHF hull and HFs whose intersection accurately approximates the input (b). We then employ MLPs (c) to encode the DHF and HFs as $R^2 \rightarrow R$ functions (on the HF only purple areas are inside the $\tilde \Omega$ parameter domain) (d).  At inference time we combine (intersect) the MLP outputs (e).}
\vspace{-.2cm}
\label{fig:overview}
\end{figure*}

\section{Overview}
\label{sec:overview}
\label{sec:define}

\paragraph{Definitions.}
Our shape representation centers around two types of {\em volumetric explicits (VEs)}: closed {\em double height field (DHF) hulls} (Fig~\ref{fig:nesi}b) and open half-spaces defined by single {\em height-fields (HFs)} (Fig~\ref{fig:nesi}c). We define both DHFs and HFs with respect to their local $x$-$y$-$z$ coordinate systems, as follows. Let $\Omega$ denote a 2D bounded domain in the $x$-$y$ plane, and let $f_a(x,y)$ and $f_b(x,y)$ be two piecewise continuous functions defined over $\Omega$, with $f_a(x,y) \geq f_b(x,y)$, $\forall (x,y)\in \Omega$; then the DHF hull is defined by $DHF = \{ (x,y,z) |  f_a(x,y) \geq z \geq f_b(x,y) \wedge (x,y) \in \Omega\}$. Here $f_a(x,y)$ and $f_b(x,y)$ are called the {\em bounding functions} of the DHF. Similarly, let $\Omega$ denote a 2D bounded domain in the $x$-$y$ plane, and let $f(x,y)$ be a piecewise continuous function defined over $\Omega$. We define $HF = \{ (x,y,z) |  z \leq f (x,y) \wedge (x,y) \in \Omega\}$. Here $f(x,y)$ is the {\em bounding function} or {\em height function} of the HF. 

When a DHF hull or an HF is assigned a general orientation, we associate it with its local coordinate system. In the notation below we use a 3D unit direction vector ${\bf d}$ to denote the $z$-axis of the local coordinate system and call ${\bf d}$ the {\em axis} of the corresponding VE.  By definition, a DHF hull is always closed, or bounded. In contrast, a HF is only half-bounded since it is unbounded in the direction of ${\bf -d}$. By these definitions, a DHF can be also viewed as the intersection of a pair of HFs with parallel but opposite axis directions.

\paragraph{ESIs and NESIs.} Our analytic {\em Explicit Surface Intersection (ESI)} and learned {\em Neural Explicit Surface Intersection (NESI)} representations use the intersection of one DHF hull $DHF({\bf d}_0)$ and zero or more HFs $HF_k({\bf d}_k)$ $(k=1,2, \dots, m)$ to approximate a given 3D object. More specifically, let $S$ denote the set of points occupied by a given 3D shape; we then approximate $S$ by the set $\tilde S = DHF({\bf d_0}) \cap_{k=1}^m HF_k ({\bf d}_k)$. By construction, the DHF hull provides a closed and tight bounding volume of $S$. Intersecting the DHF hull with the HFs further tightens this bounding volume to achieve an accurate approximation of the input. The key to our method is the observation that the vast majority of shapes %
can be well approximated using the intersection of one DHF and a very small number of HFs, with judiciously selected coordinate system axes. 

\paragraph{Computing ESIs.} Acting on the observation above, we propose an effective and efficient algorithm for computing the combination of a DHF hull and an as-small-as-possible number of HFs whose intersection accurately approximates the input shape (Sec.~\ref{sec:approximation}). As our evaluation (Sec.~\ref{sec:results}) demonstrates, on average one DHF and two HFs are amply sufficient to well approximate typical geometric shapes, and many shapes (31\% in our experiments) can be well approximated using a single DHF. %

\paragraph{Computing and Utilizing NESI.} 
We convert ESIs into a neural form by learning neural representations of the individual volumetric explicits (VEs) (Sec.~\ref{sec:learning}). We minimize the size of the learned DHF representations by leveraging the relation between their two bounding functions, and reduce the size of the individual HF encodings by only learning their surface shape in areas not well-represented by the DHF or other HFs.  
Finally, we propose efficient algorithms for performing common geometry processing tasks directly on the learned NESI representations (Sec.~\ref{sec:processing}).  NESI's  support for real-time in-out query computation enables fast ray-tracing, collision detection, and other similar tasks; at the same time, explicit surface parameterization of the individual VEs enables other tasks such as texturing and meshing.   

\paragraph{Extension to Occluded Surfaces.}
In our target applications, such as video games or VR/AR immersion, the viewer is typically located outside of rendered shapes and does not see content that is not well visible from outside; our core method targets this setting and implicitly prioritizes approximation of visible surface areas. We efficiently extend our method to approximate shapes containing fully or partially occluded surfaces by using unions of volumetric NESI explicits $\tilde S$ (see Fig.~\ref{fig:extension}, Appendix~\ref{sec:extensions}). Unless specifically indicated otherwise, all results and measurements reported and shown in the paper are generated {\em without} this extension. 

\section{ESI Computation}
\label{sec:approximation}
\label{sec:construction}
\label{sec:esi}

Given an input 3D shape $S$, represented using a triangular mesh, we seek to approximate it as an intersection of DHF hull and zero or more HFs. Since processing time and memory footprint both increase with the number of HFs, we aim to keep this number as small as possible, while maximizing approximation quality. 

Our approximations must satisfy two properties: {\em volumetric approximation} and {\em surface coverage}. The former property requires the intersection of the volumetric explicits we use to closely overlap the input shape $S$, and is critical for reliable in-out queries. The surface coverage property requires the bounding functions of the participating volumetric explicits (VEs) to jointly cover the surface. This property enables bijective piece-wise parametric representation of the outer surface, or shell, of $S$. 

To satisfy these properties, for an HF or DHF with a known $z$-axis ${\bf d}_k$, we position its $x$-$y$ plane just under the bounding sphere of $S$ along the axis direction (at the sphere-axis intersection) and define the domain $\Omega$ of the VE in this plane as the 2D region bounded by the silhouette of the object viewed along the axis direction. For the DHF, denoted by $DHF ({\bf d}_0)$, we define its bounding functions $f_a(x,y)$ and $f_b(x,y)$, over the domain $\Omega_0$, as the two depth maps of the shell $S$ when viewed along the $-{\bf d}_0 $  and ${\bf d}_0$ directions, respectively.
For any given HF, denoted by $HF_k ({\bf d}_k)$,  we define its bounding function $f_k(x,y)$ over the domain $\Omega_k$ as the depth map of the shape $S$ when viewed along the direction  ${\bf d}_k$ (see Fig~\ref{fig:nesi},~\ref{fig:overview}b).  This formulation ensures that the input shape $S$ is entirely contained inside each VE, and that each of the VE's bounding functions overlaps with the depth map of $S$ with respect to the corresponding axis. 

\begin{figure}
\includegraphics[width=0.9\linewidth]{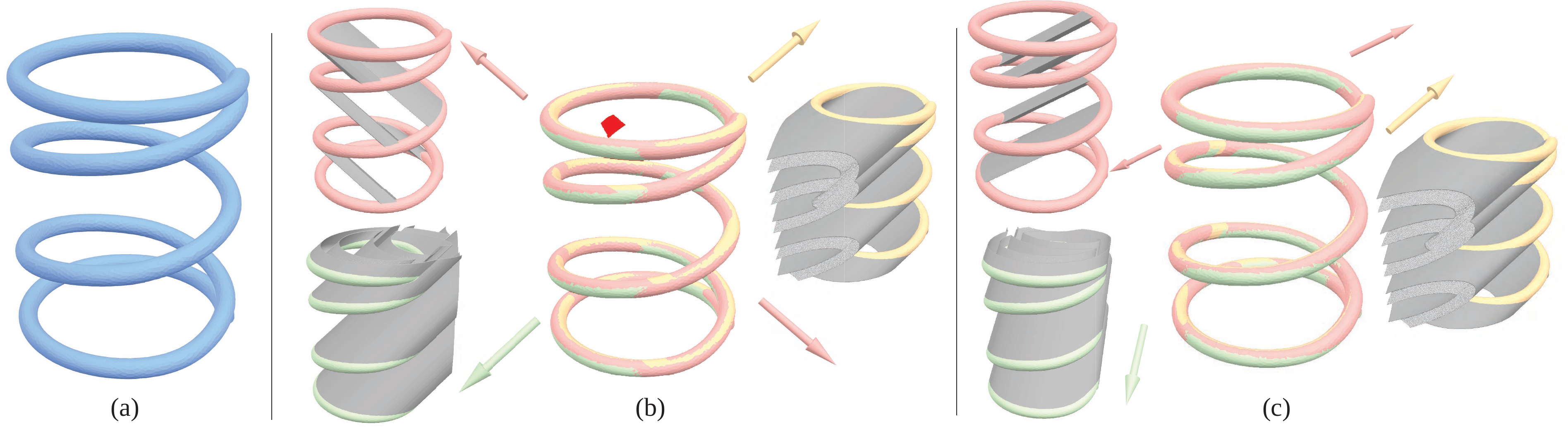}
\caption{Approximation criteria: given the inputs on the left (a), using only surface coverage as an axis selection criterion produces VE intersections that may contain extra undesirable connected components  (b, highlighted in red). Optimizing for both volumetric approximation {\em and} surface coverage produces the outputs we seek (c). }
\label{fig:chair_spiral}
\end{figure}

\begin{parWithWrapFigure}
\begin{wrapfigure}{l}{.17\columnwidth}%
\includegraphics[width=\linewidth]{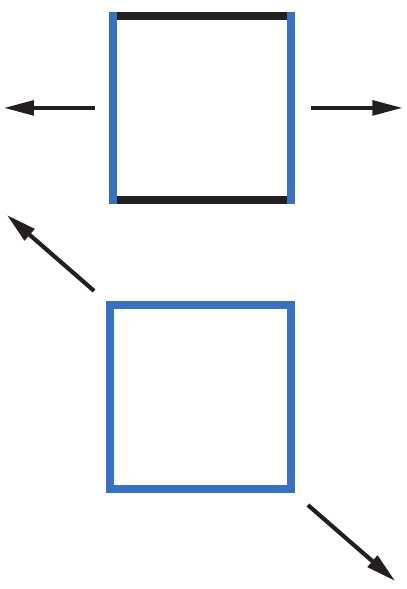}%
\end{wrapfigure}
With this definition in place, the problem of computing an optimal set of VEs can be recast as one of computing the optimal set of VE axis directions that best satisfy the two criteria above. We note that optimizing either volumetric approximation or surface coverage in isolation can produce outputs poorly suited to our needs. In the inset, the top, horizontal, DHF axis choice  produces a DHF that accurately approximates the input shape, but whose bounding functions only cover the square's sides, whereas using the diagonal axis on the bottom produces a DHF that satisfies both volumetric approximation and surface coverage.  Perhaps less intuitively, as Fig~\ref{fig:chair_spiral} shows, surface coverage does {\em not} guarantee volumetric approximation: while the choice of axes in (b) results in a set of VEs whose bounding functions cover the input shape, the intersection of the corresponding volumes contains an additional undesirable connected component. 
\end{parWithWrapFigure}

Based on these observations, we measure the quality of a given approximation $\tilde S$, defined in terms of the participating DHF and HF axes, as   

\begin{equation}
E(\tilde S) = \operatorname{Dist}(S, \tilde S) + \min(\operatorname{Cov}(\tilde{S}) (\operatorname{Dist}(S,\tilde S) + \varepsilon_{1}), \varepsilon_{2})
\label{eq:esi}
\end{equation}

Here $\operatorname{Dist}(S,\tilde S)$ measures the bidirectional $L_{1}$ Hausdorff, or closest, distance between $S$ and $\tilde S$; and $\operatorname{Cov}(\tilde S)$ measures the quality of the surface coverage of $S$ provided by the bounding functions of the participating VEs:

\begin{equation}
\operatorname{Cov}(\tilde{S}) = \left ( \frac{ \operatorname{Area}(\tilde S) - \operatorname{Area} (\cup f_k)} {\operatorname{Area} (\tilde S)} \right ) 
\end{equation}

The combined loss function balances the two criteria while prioritising volumetric approximation. The set of variables we operate on is the number of HFs and the axis directions of the DHF and participating HFs; we seek to efficiently compute the combination of axis directions that minimizes $E(\tilde S)$.  
Unfortunately, even just computing $E(\tilde S)$ for a given set of VE axes is highly time consuming, as it requires computing the geometry of the participating VEs, computing their Boolean intersection, and then finally computing the distance between this intersection and the input shape. To make this optimization tractable, we rely on a discrete optimization process that leverages the unique geometric properties of our DHF and HF shapes, and an effective branch-and-bound scheme that exploits our problem setup.

We first discretize $E(\tilde S)$ by sampling both $S$ and $\tilde S$ uniformly and densely, producing sets of points $P$ and $\tilde P$. $\operatorname{Dist}$ is then evaluated point-to-point on these two sets (a.k.a. $L_1$ chamfer distance).  We avoid explicit computation of the bounding functions $f_k$. Instead, for each point $\tilde p \in \tilde P$, we estimate its likelihood of being on $f_k$ based on its visibility $v(\tilde p,d_k)$ along the axis $d_k$ and the angle between $\tilde p$'s normal $\hat{n}_{\tilde{p}}$ and the axis direction.  We set $v(\tilde p,d_k)$ to 1 if the point is visible (the ray from $\tilde p$ along ${\bf d}_k$ does not intersect the surface) and 0 otherwise. We set

\begin{equation}
\operatorname{On}(\tilde p, d_k )=\max ((1- v(\tilde p,d_k)), T( \cos^{-1} (\hat{n}_{\tilde p} \cdot {\bf d}_k))) 
\end{equation}

\begin{parWithWrapFigure}
	\begin{wrapfigure}{l}{.2\columnwidth}%
		\includegraphics[width=\linewidth]{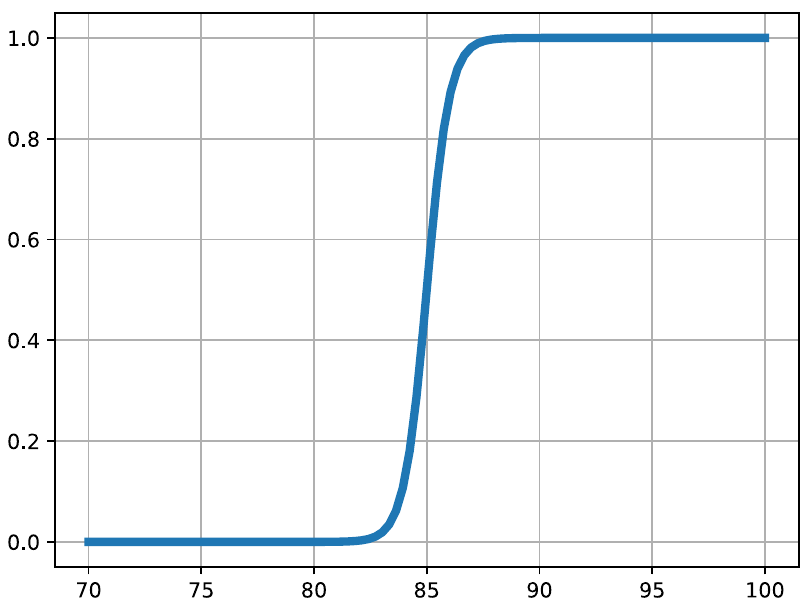}%
	\end{wrapfigure}
Here $T$ is the shifted and scaled $\tanh$ function shown in the inset, chosen so that $T$ is $0$ if $\hat{n}_{\tilde p}$ is well-aligned with ${\bf d}_{k}$ (the angle between them is significantly below $90^{\circ}$) and increasing to $1$ as the angle approaches or exceeds $90^{\circ}$.  
\end{parWithWrapFigure}

We then define $\operatorname{Cov}$ as 
\begin{equation}
\operatorname{Cov}(\tilde{S})=\frac{1}{|\tilde P|}\sum_{\tilde p=1}^{|\tilde P|} \min_{k} \operatorname{On}(\tilde p, d_k )
\end{equation}

Even with this discretization in place, computing $E(\tilde S)$ takes a non-trivial amount of time, as it requires sampling points on $\tilde S$ which is not explicitly defined. We therefore seek to minimize the number of $E(\tilde S)$ evaluations.

To this end, rather than optimizing over an infinite set of possible axis directions, we use a finite set of well-distributed potential axis direction samples in our implementation (50 for DHF, and 80 for HFs). Since each direction vector corresponds to a point on the unit sphere, we evenly pick HF and DHF axis candidates by sampling these points on the unit sphere $S^2$ using spherical Fibonacci sampling \cite{keinert2015fibonacci}. Since many objects encountered in practice are axis-aligned, we augment our sampled candidate directions with the three major axes (both directions).

As we expect approximation quality to improve as more HFs are added, we compute solutions incrementally for each possible HF count $m$, starting with $m=0$ (i.e a DHF only); we terminate the process only once adding an extra HF fails to improve accuracy, or a maximal number of HFs is reached (we cap this number at 3; see Sec. ~\ref{sec:results} for validation of this choice.) Even with $m=3$, however, the space of all possible axis combinations has 25.6 million combinations ($50\cdot80^3$), necessitating both a highly efficient strategy to evaluate $E(\tilde S)$ and a robust search method that minimizes the number of evaluations required. 

To achieve this speedup, we first recall that the term $\operatorname{Dist}$ is bidirectional, and can be written as the sum of two nonnegative, one-directional Chamfer distances (denoted $\operatorname{CD}$): the Chamfer distance between the candidate ESI and the mesh, and the Chamfer distance between the mesh and the ESI:

\begin{equation}
\operatorname{Dist}(S,\tilde S) = \operatorname{CD}(S,\tilde S) + \operatorname{CD}(\tilde S, S)
\end{equation}

We observe that we can sample points on our DHF and HFs by leveraging simple-to-evaluate in-out queries: a point is inside an HF (defined as above) if and only if a ray originating at the point and emanating along the HF direction axis intersects the input surface $S$, and a point is inside a DHF if and only if rays emanating along both positive and negative axis directions intersect $S$. We use this observation in a ray-casting framework to robustly compute points on the surface of all potential DHF and HF volumes. Notably, this computation is done as a pre-process, once for each potential DHF/HF axis. We further observe that an immediate consequence of this framework is that computing $\operatorname{CD}(S, \tilde S)$ is {\em much} faster than computing the inverse distance $\operatorname{CD}(\tilde S, S)$, as it can be expressed in terms of distances between precomputed sample points on $S$ and the participating VEs. Furthermore, as $\operatorname{CD}$ is strictly nonnegative, we can use $\operatorname{CD}(S,\tilde S)$ as a lower bound to quickly reject candidate VE combinations against the best known solution. We therefore search for VE candidates in parallel, and for each VE combination we only proceed to evaluate $E(\tilde S)$ in full if $\operatorname{CD}(S,\tilde S)$ is lower than the best quality score encountered so far. This branch-and-bound strategy reduces the number of full evaluations for $m=2,3$ by $98\%$ on average. For an additional speed up, before testing points and rays against $S$ we reject sample points and ray directions that do not intersect the convex hull of $S$.

Overall our optimized ESI computation takes 3 minutes on average on a 16-core Intel Xeon Gold 6130 CPU (an average of 2 minutes of preprocessing time, and 1 minute or less for axis selection), and up to 12 minutes on complex models like the {\em happy buddha} (Fig. ~\ref{fig:teaser}); 9 minutes preprocessing time, 3 minutes axis selection). %

\section{Learning NESIs}
\label{sec:learning}
\label{sec:nesi}

Once ESI is computed, we encode the bounding functions $f_a$, $f_b$, and $f_k$ of its participating VEs in a compact neural form as a set of MLPs. We now describe how to determine the domain $\Omega$ of each bounding function; the loss functions for training; the point sampling strategy in each domain used for evaluating the loss functions; and the overall neural network architecture. 

To operate on a bounded numerical range, we define each explicit in its local axis-aligned coordinate system and restrict it to a $[-1,1]^3$ bounding box. We normalize our 3D shapes $S$ to be strictly inside this box by scaling them to be inside of $[-\frac{1}{1.1}, \frac{1}{1.1}]^3$. All our learned functions are thus defined over $D_k = [-1,1]^2$.

\paragraph{DHF Hull Domain Sampling.}
A fully-supervised training of a network that 
reproduces the DHF bounding functions $f_a$ and $f_b$ requires first obtaining sample points $(x_i,y_i) \in \Omega_0$, together with the ground-truth function values at these sample points $(x_i,y_i)$ for supervision. 
To generate sample points for learning $f_a(x,y)$ and $f_b(x,y)$, we first uniformly sample a dense set of points $Q=\{q_i\}$, where $q_i=(x_i,y_i,z_i)$, on the surface of the shape $S$. We then filter these points to find a set of samples $Q_a$ that well approximates $f_a(x,y)$. We set $Q_a=Q$, then shoot rays from each point $q_i \in Q_a$ along ${\bf d}_0$, and remove $q_i$ from $Q_a$ if the ray intersects $S$.
We generate $Q_b$ in a similar way, but using rays along $-{\bf d}_0$. 
We project the sample points in $Q_a \bigcup Q_b$  onto the $x$-$y$ plane to obtain $P_0=\{p_i\}$, where $p_i=(x_i,y_i)$. 
By construction, the projected sample points $p_i=(x_i,y_i)$ all lie within $\Omega_0$ and provide a dense covering of the domain.
For each $p \in P$, we compute the corresponding function values $f_a(p)$ and $f_b(p)$ (the surface intersections that are furthest apart along the DHF axis), as well as the surface normals $n_a(p), n_b(p)$ at these intersections. 
 
In addition to the height functions themselves, we must also encode or store the domain $\Omega$ of the bounding functions for the DHF and HFs. For the DHF hull, we observe that we can encode this domain intrinsically by requiring that $f_b(x,y) < f_a(x,y)$ for all $(x,y) \in \Omega_0$, and $f_b(x,y)$ $\geq$ $f_a(x,y)$ for all $(x,y) \notin \Omega_0$. To obtain sample coverage for $D_0 \setminus \Omega_0$, we sample points in $D_0$,  then shoot rays from each point along both ${\bf d}_0$ and $-{\bf d}_0$, discarding points if the ray intersects $S$. The remaining points form our $\overline{P_0}$ set.

\paragraph{HF Domain Sampling.}
We generate sample points for learning the height functions $f_k$ of each HF volume $HF_k({\bf d}_k)$ ($k=1,2\dots, m$) using a similar process, with one major difference (Fig~\ref{fig:nesi}de). 
We observe that each explicit provides two types of information about the approximated shape - the outline of its visual hull ($\Omega_k$) and the shape geometry inside this outline.   
Our explicits often cover overlapping regions on the input shape (Fig~\ref{fig:overview}b); encoding the geometry in these regions more than once introduces unnecessary redundancy, wasting network capacity. To avoid such redundancy we seek to restrict the parameter domain for which we store the geometry (height values) of each additional {\em learned} HF to only span those surface regions on the input shape $S$ that have not been 
well-covered by the combination of the DHF and any previous HFs. We denote surface regions which are covered by $HF_k({\bf d}_k)$, but {\em not well covered} by prior explicits, as $S_k$ (in Fig~\ref{fig:overview}c the purple regions on the HF correspond to areas not well covered by the DHF). A point $q$ on $S$ is in $S_k$ if  the ray from $q$ along ${\bf d}_k$ does not intersect the surface {\em and if one} of the following three conditions holds:

\smallskip
\noindent
(1) A ray from $q$ along any of the directions $\{{\bf d}_0, -{\bf d}_0, {\bf d}_1,\dots,{\bf d}_{k-1}\}$ intersects $S$; in other words, $q$ is not on the surface of the DHF or any preceding HFs and hence is not represented by them,

\noindent
(2) $q$ is on the surface of the $DHF$ or any previous $HF_i$ but is nearly a {\em grazing point} of $S$ with respect to the axis of this previous HF or DHF; that is, the angle between the surface normal vector at $q$ and the corresponding axis ${\bf d}_i$ is larger than a minimum threshold ($70^\circ$ in our implementation). The rationale behind this condition is twofold. First, we seek an approximation which supports low distortion parameterization, thus we prefer surface regions to be approximated using HFs which better align with their normal, and to specifically avoid high parametric distortion. Moreover, small errors in height function approximation in grazing regions can result in large approximation error in $\mathbb{R}^3$. 

\noindent
(3) There exists a point $a$ on the ray from $q$ along ${\bf d}_k$ that is invisible from all previous directions $\{{\bf d}_0, -{\bf d}_0, {\bf d}_1,\dots,{\bf d}_{k-1}\}$. This condition is critical for approximating the interior of large voids or concavities which are not captured by previous explicits. 
\smallskip

Formally, the domain $\Omega_k$ of a height field $HF_k$ is defined the same way as for the $DHF$, delineating the outline of the HF's visual hull. 
However, for height learning purposes, we restrict $HF_k$  to only cover the region $S_k$ not properly covered by the $DHF$ or any preceding $HF$s and focus on the subdomain of $\Omega_k$ given by the projection of $S_k$ in the direction of ${\bf d}_k$ onto the $x$-$y$ plane; we denote it as $\tilde \Omega_k$. We note that with these restricted domains $\tilde \Omega_k$ in place, one can recast the Boolean definition of NESI using subtraction instead of intersection (see Sec,~\ref{sec:discuss}). 

With these criteria in place, we compute three sets of training samples. We first form $Q_k$, the subset of samples on $S$ such that a ray from $q \in Q_k$ along ${\bf d}_k$ does not intersect $S$. We then compute the subset of points $Q'_k \in S^k$ by evaluating if they satisfy one of the criteria above. We project these points to the $x$-$y$ plane of $HF_k$ forming $P_k$ and $P'_k$, and form $\overline{P_k}$ using the same process as for $\overline{P_0}$.

In addition to the height functions themselves, we must also encode the domains $\Omega_k$  and   $\tilde \Omega_k$ of the bounding functions for each HF.  The former defines the visual hull of the shape along the axis $d_k$ while the later encodes the domain within which we want to encode the HF geometry. We encode $\tilde \Omega_k$ implicitly by forcing the height values at points inside $P_k \setminus P'_k$ to be above the surface. We use an explicit binary mask $\delta_k$ to specify the domain $\Omega_k$ and  encode each HF as an MLP that returns both the bounding function $f_k$ and the mask $\delta_k$; we observe that this mask can be constructed to require significantly less network capacity than $f_k$ itself.

\begin{parWithWrapFigure}
\begin{wrapfigure}{l}{.35\columnwidth}
\includegraphics[width=\linewidth]{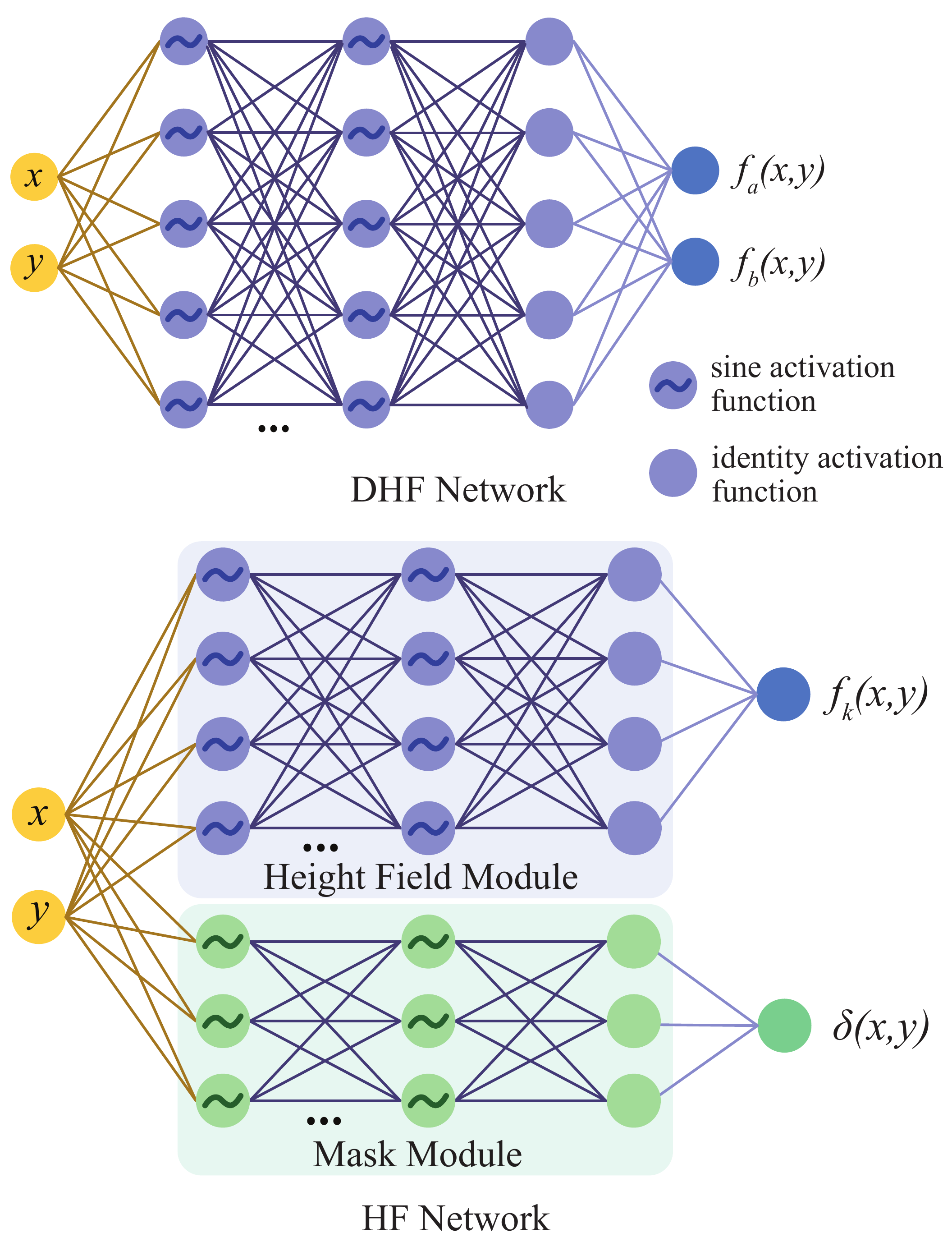}
\end{wrapfigure}
\paragraph{Network Architecture.}
We adopt the SIREN architecture \cite{siren} for our MLP network. Our network takes 2D locations $(x,y)$ as input.
For the DHF, the network outputs two values, each corresponding to the height values of the two bounding functions $f_a(x,y)$ and $f_b(x,y)$ of each side of the DHF (inset, top). 
We encode the height and mask of each HF as two separate MLP networks (inset, bottom). The height function module outputs a single height value $f_k(x,y)$ for all points in the restricted domain ${\tilde \Omega}_k$; inside ${\Omega \setminus \tilde \Omega}_k$, the network is trained to output a height value that is greater than ground truth (i.e. $f_k(x,y)>f_{gt}(x,y), \forall (x,y) \in P'_k$). The domain of ${\Omega}_k$ is encoded as a binary indicator function $\delta(x,y)$ by a separate compact MLP.
Both networks take 2D coordinates $(x,y)$ as input, and pass them through $n$ hidden layers with sine activation functions between adjacent ones. For DHFs, we output two height field values to bound the finite volume of the target shape. For HFs and their masks, we use a smaller SIREN network to infer height field values and a very compact network to infer an indicator $\delta$ which indicates whether a point is inside or outside the projected shape region $\Omega$.
The final encoded NESI representation consists of $2m+1$ MLPs: 1 MLP, the largest one, for the DHF; $m$ smaller ones for the HFs; and $m$ tiny MLPs for the HF masks. 
\end{parWithWrapFigure}

\paragraph{Objective Functions.}
To learn the neural encodings of the DHF hull and our $k$ HF volumes we minimize the following loss functions, defined in the local coordinate frame of each explicit volume. 

Our DHF loss is defined as
\begin{equation}
	\loss^{D}=\mathcal{L}_{Height}^{D}+\loss_{Domain}^{D}+\alpha_{Normal}\mathcal{L}_{Normal}^{D}  \label{eq:loss_dhf}
\end{equation}
where the first terms encode the heights of the explicit surfaces; the second delineates $\Omega_0$; and the third term encodes the explicit surface normals: 
\begin{align}
	\loss_{Height}^{D}=&\frac{1}{|P_0|}\sum_{\point \in {P_0}}\left(|f_a(\point)-f_a^{gt}(\point)|+|f_b(\point)-f_b^{gt}(\point)|\right) \\[-2pt]
	\loss_{Domain}^{D}=&\frac{1}{|\bar P_0|}\sum_{\point \in \bar P_0 }max(f_a(\point)-f_b(\point)+\epsilon, 0) \\[-2pt]
	\loss_{Normal}^{D}=&\sum_{i \in a,b}\frac{1}{|P_0|}\sum_{\point \in P_0}(1-\frac{n_i(\point)\cdot n_i^{gt}(\point)}{\max(\|n_i(\point)\|_2\cdot\|n_i^{gt}(\point)\|_2,\epsilon)}) \label{eq:normal_dhf}
\end{align}
Here $f_i(\point)$ and $f_i^{gt}(\point)$ are the learned and input function values, $n_i^{gt}(\point)$ is the input surface normal at $\point$, and $n_i(\point)$ is the normal computed by 
backpropagating the network that learns $f_i$. The weight $\alpha_{Normal}$ is used to suppress the normal loss at the beginning of training, then gradually increase it after a certain number of iterations. We set this weight as $\alpha = 0.5\tanh((i - 4000)/10)+ 0.5$, where $i$ is the current iteration.

The loss function for the $k$th HF is defined as  
\begin{equation}
	\loss^{H}(k)=\mathcal{L}_{Height}^{H}(k)+ \loss_{Domain}^{H}(k) + \loss_{Mask}(k)+\alpha_{Normal}\mathcal{L}_{Normal}^{H}(k) \label{eq:loss_hf}
\end{equation}
where $\loss_{Domain}^{H}$ codifies the function behavior across  $\Omega_k\setminus{\tilde \Omega}_k$, and $\loss_{Mask}$ defines the mask used to delineate $\Omega_k$.

\begin{align}
	\loss_{Height}^{H}(k)=&\frac{1}{|P'_k|}\sum_{\point \in {P'_k }}|f_k(\point)-f_k^{gt}(\point)| \\[-1pt]
	\loss_{Domain}^{H}(k) =& \frac{1}{| P_k \setminus P'_k|}\sum_{\point \in P_k \setminus P'_k}max\left(\left(f_k^{gt}(\point)-f_k(\point)\right)+\epsilon, 0\right) \\[-1pt]
	\loss_{Mask}=& \frac{1}{|\tilde P_k|}\sum_{\point \in \tilde P_k}\bce(\delta_i(\point),\delta_i^{gt}(\point))\\[-1pt]
	\loss_{Normal}^{H}=&\frac{1}{|P'_k|}\sum_{\point \in P'_k}\left (1-\frac{n_k(\point)\cdot n_k^{gt}(\point)}{\max(\|n_k(\point)\|_2\cdot\|n_k^{gt}(\point)\|_2,\epsilon ) }\right ) \label{eq:normal_hf}
\end{align}

Here $\bce$ is the binary cross entropy (BCE) loss \cite{good1952rational}, and $\tilde P_k = P_k \cup \bar P_k $ is the collection of all previously computed sample points for this HF.

\section{Using NESI Representations}
\label{sec:use}
\label{sec:processing}
\label{sec:inference}

Finally, we show how NESI can be used as either an implicit surface or as a piecewise parametric representation.

\paragraph{NESI Occupancy Function.}
NESI trivially supports in-out occupancy tests. For a given 3D point $\point$, we simply check whether it is inside or outside of the DHF and any HFs: we first convert $\point$ to the local coordinates $(x,y)$ of each DHF and HF; for HFs, we check if the $(x,y)$ in this coordinate system is inside the domain using the mask; and finally we compare the associated height $z$ to the predicted height values from the network to determine if it is inside or outside based on the definition of explicits. The point is inside the NESI if and only if it is inside all its explicits.

\paragraph{NESI as a Parametric Representation.} NESI's parameter domain, or atlas, consists of $\Omega_0$, used twice for the DHF hull top and bottom; and $\tilde \Omega_k$ for each HF.  To map a surface point $\point$ to the atlas, we locate the explict whose surface it is closest to and then use the $(x,y)$ coordinates of the point in the coordinate system of this explicit as its $(u,v)$ parameters.

\section{Results}
\label{sec:results}

\begin{figure}
	\includegraphics[width=\linewidth]{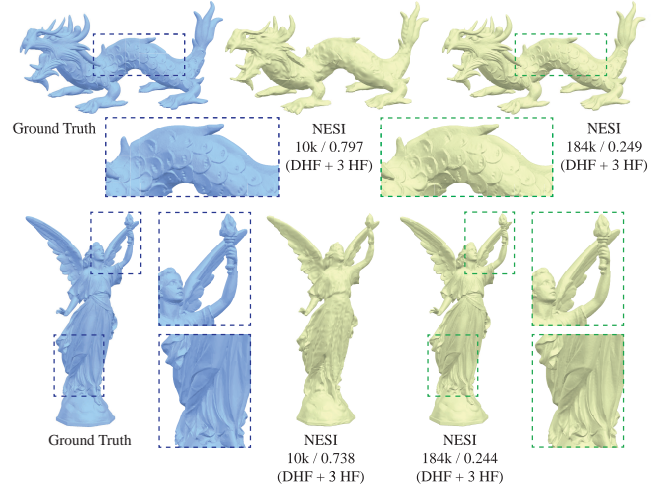}
	\caption{NESI approximations of the {\em lucy} and {\em xyzdragon models} at low (10k) and extra high (180k) parameter counts. At low parameter counts, NESI captures the core properties of the inputs; at high parameter counts, even very fine details such as {\em lucy}'s torch and the scales of the dragon are accurately represented.}    
	\label{fig:extreme}
\end{figure}

We evaluate our analytic (ESI) and neural (NESI) representations qualitatively, via visual inspection; and quantitatively, by measuring the distance (chamfer $\Lone$) between the inputs and our analytic and learned outputs. We compare our results against an extensive list of alternatives, and ablate different algorithmic choices as discussed below. Finally, we showcase applications of using NESI for different graphics applications, discuss its limitations, and propose extensions addressing those.  Throughout the paper we showcase 57 representative NESI outputs highlighting their high visual quality, which remains high even at low parameter counts (e.g. Fig. ~\ref{fig:extreme},~\ref{fig:qual_all},~\ref{fig:siren_qual}, ~\ref{fig:ngf}).  All renderings of both our and alternative results were generated via our raytracing code and colored using a flat shading scheme. Our raytracer uses NESI's (and ESI's) ability to instantaneously and robustly evaluate in-out queries (Sec~\ref{sec:inference}). 
See the appendix and supplementary material for input sourcing details, implementation details, galleries of input and output visuals, and additional details of the evaluations below. 

\paragraph{Evaluating ESI.}
To evaluate the premise behind our ESI and NESI representations, and their robustness, we seek to answer two questions: first, how accurately can the ESIs computed using our method (Sec~\ref{sec:esi}) approximate typical content rendered by consumer facing graphics applications; and secondly, how many VEs are required to approximate such shapes to a desired accuracy using our method?  
  
To answer these questions, we assembled a corpus of 320 diverse inputs representative of the types of geometries rendered by the applications we target. Our corpus included the Thingi32 \cite{nglod,thingi10k} (32 shapes) and DHFSlicer~\cite{yang2020dhfslicer}(25 shapes) datasets representative of related prior work; non-trivial random shapes from the the ABC~\cite{abcdataset} dataset of CAD models (40 shapes); the dataset of  \mylesdata (98 shapes commonly used in computer graphics); 122 additional random inputs from Thingi10K \cite{thingi10k}; and complex, canonical, scanned shapes from the Stanford 3D Scanning Repository \cite{StanfordScanRep} ({\em david, dragon,} and {\em thai statue}). These inputs span both CAD  and organic content, include highly complex shapes ({\em david, thai statue,} and {\em lucy}), as well as shapes with high genus, non-manifold geometry, and other artifacts.  %

To answer the first question, we used the method in Sec~\ref{sec:esi} to generate ESI approximations of these shapes using a fixed number of HFs (ranging from 0 to 4). We then measured the chamfer distance between the ground truth shapes and these ESI approximations (\Tab{viewdir}, all distances multiplied by 1000 and measured relative to the input bounding box diagonal). As the numbers show, even for a single DHF with no additional HFs, the approximation quality is often already very good. The chamfer distance decreases with additional HFs, but the amount of improvement tapers out, motivating our cutoff of using no more than 3 HFs. To provide some context to the numbers reported, we recall that while the Hausdorff distance between a surface and itself is zero, chamfer distance is an approximation of Hausdorff distance and is measured by computing the distance between point clouds sampled on the two surfaces. In particular, chamfer distance depends on sampling density - accuracy increases as point cloud size grows. Chamfer distance between a surface and itself, measured using two different clouds sampled from the same surface, will never be strictly zero, but is expected to decrease as sampling density increases. To obtain high accuracy distances we use very large clouds (5M points). Even with this high density, the baseline chamfer from our inputs to themselves is $0.219$.  When using up to 3 HFs our ESI approximation quality (average chamfer $0.234$) is within $0.015$ of these values; in short, our chamfer distance is extremely close to the chamfer distance between our input meshes and {\em themselves.}  
These numbers confirm the main insight behind our method: {\em typical 3D geometries used in computer graphics applications can be well approximated by intersecting a small number of judiciously selected volumetric explicits.} The ESI to input distances can be thought of as a lower bound on the approximation quality provided by the corresponding learned NESI approximations. 

\begin{table}
\begin{center}
	\tiny
	\begin{tabular}{l r c c c c c}
	\toprule
	Dataset & DHF only & DH + 1 HF & DHF + 2 HFs & DHF + 3 HFs  & DHF + 4 HFs & \# VEs used \\
	\midrule
	\cite{Myles16}  & 0.44 (1.10) & 0.24 (0.49) & 0.22 (0.28) & 0.22 (0.24) & 0.21 (0.23) & 3 (2.82)\\
	Thingi10k  & 0.25 (0.73) & 0.23 (0.33) & 0.23 (0.26) & 0.23 (0.24) & 0.23 (0.23) & 1 (2.03)\\
	Thingi32 & 0.45 (0.69) & 0.26 (0.30) & 0.23 (0.24) & 0.23 (0.23) & 0.22 (0.23) & 3 (3.16) \\
	abc & 0.36 (1.64) & 0.28 (0.67) & 0.26 (0.30) & 0.26 (0.27) & 0.26 (0.27) & 2 (2.15) \\
	Other & 0.40 (0.62) & 0.26 (0.33) & 0.22 (0.22) & 0.22 (0.22)  & 0.22 (0.22) & 3 (2.89) \\
	\midrule
	Overall & 0.33 (0.94) & 0.25 (0.41) & 0.23 (0.27) & 0.22 (0.24) & 0.22 (0.23) & 3 (2.48) \\
	\bottomrule
	\end{tabular}
\end{center}
\caption{Median and average (in brackets) Chamfer distances between input shapes and analytic intersections of volumetric explicits best approximating them, for different numbers of HFs across the different data sources and overall. The average lower bound on chamfer distances between input shapes and themselves is 0.22, representing the maximum accuracy achievable with this metric; our average and median ESI chamfer distances meet or are close to this lower bound in many cases, representing extreme accuracy compared to ground truth.}
\label{tab:viewdir}
\end{table}

To answer the second question, we computed the ESIs of these shapes, this time letting our method determine the output number of HFs used automatically. In our experiments, we did not use an error tolerance to determine the number of HFs needed, and added additional HFs if doing so reduced the Chamfer distance by {\em any} amount.  In practice we expect users to specify a tolerance for the accuracy they need, limiting the number of HFs generated further.
We then measured the number of HFs generated for each input (\Tab{viewdir}, last column).  Across the set of 320 inputs tested, 98 inputs only required a single DHF (31\%); 59 used 1 HF (18\%); 75 used 2HFs (23\%); and 88 used three HFs in addition to a DHF (28\%). This experiment highlights the efficacy of our approach - the median input requires just 1 DHF and 2 HFs to produce accurate approximations. 

\paragraph{Evaluating NESI.}
We evaluate our neural NESI representation by learning 400 NESI models using diverse parameter counts and ground truth input shapes. 
Specifically, we use a subset of the dataset above containing the Thingi32 \cite{nglod,thingi10k} (32 shapes) and DHFSlicer~\cite{yang2020dhfslicer} (25 shapes) datasets, the subset of the ABC~\cite{abcdataset} dataset (40 shapes), and the 3 models from \cite{StanfordScanRep} for a corpus of 100 input shapes total. These inputs are representative of the type of content we target, as well as shapes processed by state-of-the-art neural representation and compression methods.  %

We approximate each input using NESI neural encoding using 4 levels of detail, or DHF/HF network parameter counts, resulting in a total of 400 encoded NESIs.
Since we target consumer facing applications, such as streaming/rendering on low memory devices, we focus most experiments on learning models using low to medium parameter counts. At the lowest level we encode each input using under 10K parameters; at the finest level, we use 50K to 70K parameters to encode each input. We note that final per-shape parameter count depends both on the specified DHF/HF network parameter counts and the number of HFs used to encode the shape. We inspect the results both visually and quantitatively.  Tab.~\ref{tab:netwidth} reports the average Chamfer $L_1$ distances between NESI approximation and input shapes for different levels of detail (all distances multiplied by 1000, measured relative to the input bounding box diagonal).  As the table and visuals (e.g. Fig~\ref{fig:extreme} (middle)) show, even at very low parameter counts NESI captures the core features of the input shapes. As desired, distance decreases as parameter count increases. We note that the average distance between NESI models trained with the largest parameter count settings and corresponding inputs (0.28) is very close to the distance between these inputs and their respective ESI approximations (0.25). This highlights the effectiveness of our neural encoding. 

\begin{table}
\begin{center}
	\tiny
	\begin{tabular}{l c c c c }
	Dataset & LOD 1 & LOD 2 & LOD 3 & LOD 4 \\
	\midrule
	ABC & 0.44 & 0.35 & 0.31 & 0.30 \\
	Thingi32  & 0.54 & 0.37 & 0.30 & 0.28 \\
	Others & 0.54 & 0.38 & 0.29 & 0.27 \\
	\midrule
	All & 0.50 & 0.36 & 0.30 & 0.28 \\
	\end{tabular}
\end{center}
\caption{Average NESI chamfer distances across different (ascending) levels of detail (parameter counts, see sup for exact settings). As desired NESI accuracy improves as parameter count increases.}
\label{tab:netwidth}
\end{table}

 To evaluate NESI's performance at higher parameter counts we additionally trained it on the {\em lucy} and {\em xyzdragon} inputs with 120K and 180K parameters each. Fig.~\ref{fig:extreme} shows the results at 180K. As the figure shows we accurately capture fine details such as the dragon's scales or the fine geometry on the dress and torch of {\em lucy}. The Chamfer distances for these experiments were: for {\em lucy}, 0.27 and 0.24 respectively; for {\em xyzdragon}, 0.26 and 0.25. For comparison, for Lucy the chamfer distance between its ESI and the input is 0.22, and between two point clouds sampled on the {\em lucy} input it was 0.19; for {\em xyzdragon}, these numbers were 0.19 and 0.17 respectively.  
These measurements confirm that as parameter count increases, our accuracy approaches that of ESI, which in turn accurately approximates the original input. 

\paragraph{Applications.} We demonstrate the versatility of NESI representations by leveraging the manipulation modes they support (implicit and parametric) for different classical geometry processing tasks. We use the implicit access mode for fast in-out queries for raytracing (used throughout the paper), and use the parametric access for texture mapping (Fig. ~\ref{fig:applications}, top) and meshing (Fig. ~\ref{fig:applications}, bottom); see Appendix for implementation details. 

\begin{figure}
	\includegraphics[width=\linewidth]{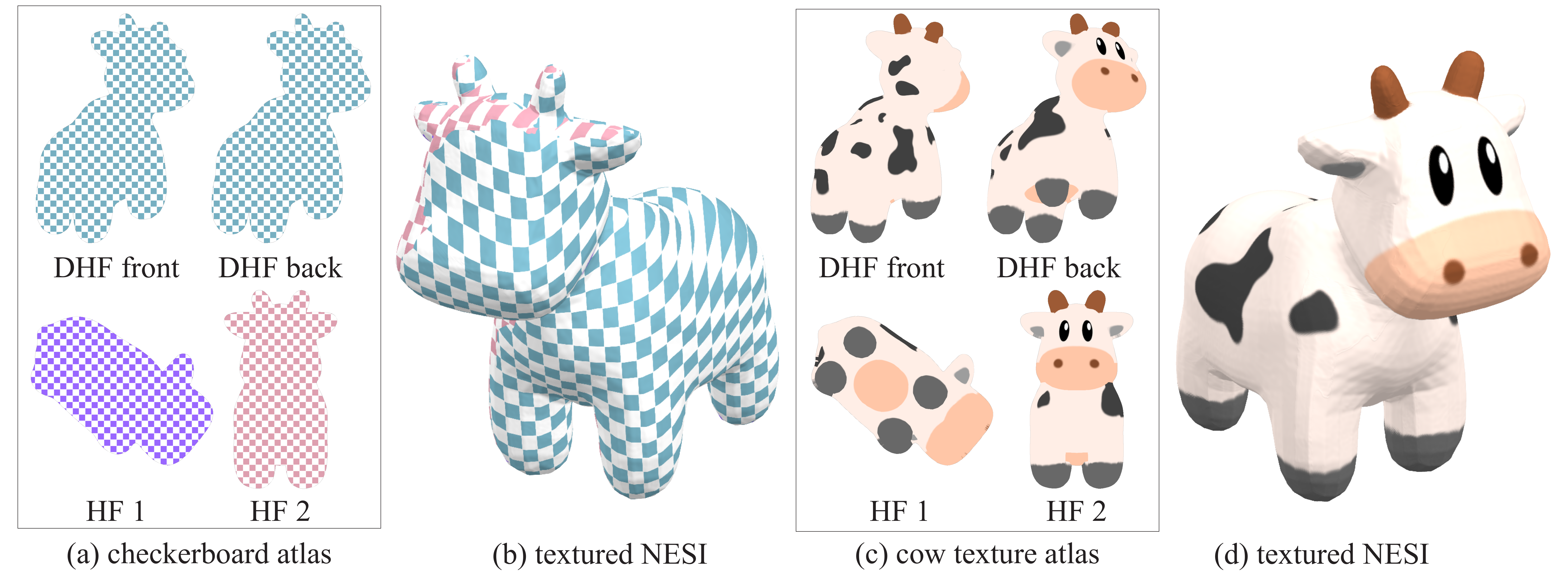}\\
	\includegraphics[width=\linewidth]{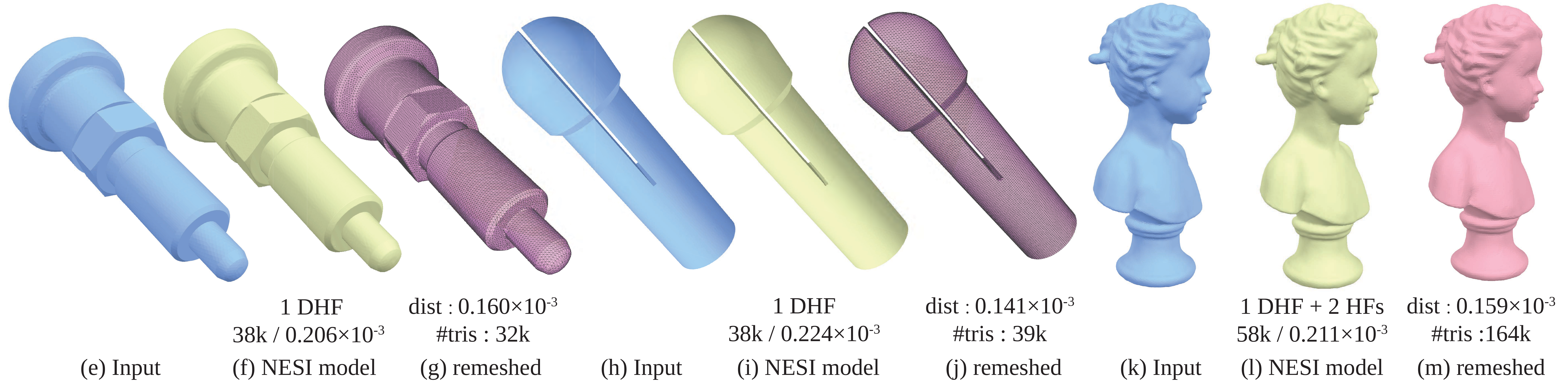}
	\caption{Applications of NESI's explicit parameterization: (Top) A learned NESI model (b,d) textured with a standard checkerboard texture (a,b)  and texture transferred from original mesh (c,d).  (Bottom) NESI models (green) learned from input meshes (blue) and then meshed using our parameterization-based method (purple); as reflected by the Hausdorff distance, the output meshes well approximate the inputs.}
	\label{fig:applications}
\end{figure}

\subsection{Comparative Evaluations}
\label{sec:compare}
\label{sec:exp}
\label{sec:exp_quant}

\begin{figure*}
	\includegraphics[width=0.9\linewidth]{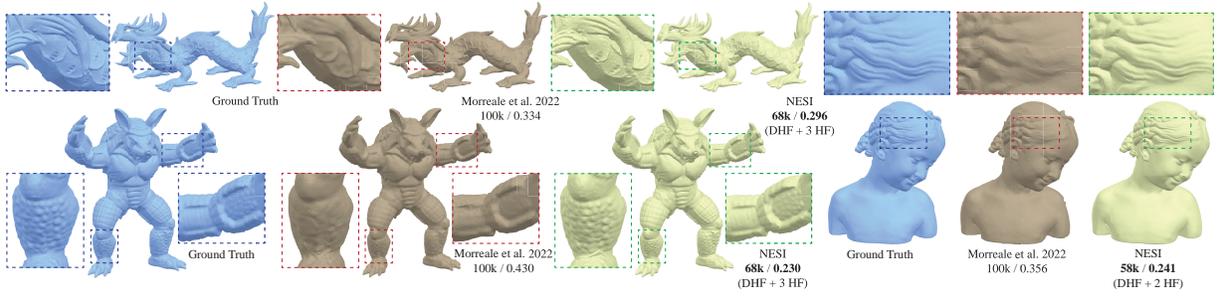}
	\caption{NESI outputs trained using 58K to 68K parameters provide a better, more detailed, approximation of the inputs than the neural models of \cite{Morreale2022NCS} which use much higher parameter counts (100K).}
	\label{fig:morreale22}
\end{figure*}

We compare our learned NESI outputs against an extensive list of representative alternatives whose authors provide either outputs or code to compare against. For fairness and consistency we use raytracing to render all outputs, and use the same point cloud sampling strategy to compute $L_1$ Chamfer distance for all outputs. (Prior works may use different sample counts and metrics in their reporting (e.g. \cite{spheres} and Sivarim ~\etal~\shortcite{ngf} report the square of $L_2$ Chamfer distances.)

We compare against the state of the art parametric neural representation method of \cite{Morreale2022NCS} by training our method on 4 example inputs they show and provide outputs for, and compare our outputs to theirs (Fig~\ref{fig:morreale22}) (the remaining 3 outputs they show are open surfaces not amenable to volumetric representation). While their respective outputs use 100K parameters, ours use only 48K to 68K parameters. Despite the lower parameter count, all of our outputs more accurately approximate the input shapes (average chamfer distance 0.33 across their outputs; 0.24 for ours): the distance for armadillo \cite{Morreale2022NCS} was 0.43, ours was 0.23 (68K); for dragon their distance is 0.33, ours is 0.30 (68K); for bimba theirs is 0.36 and ours 0.24 (58K); and for seahorse their distance is 0.22 and ours 0.18 (48K) (notably the numbers reported for their results in their paper are even larger as they used smaller point clouds to sample the input and outputs). As Fig. ~\ref{fig:morreale22} shows, our results retain significantly more visual details.
 
\begin{figure*}
	\includegraphics[width=.9\linewidth]{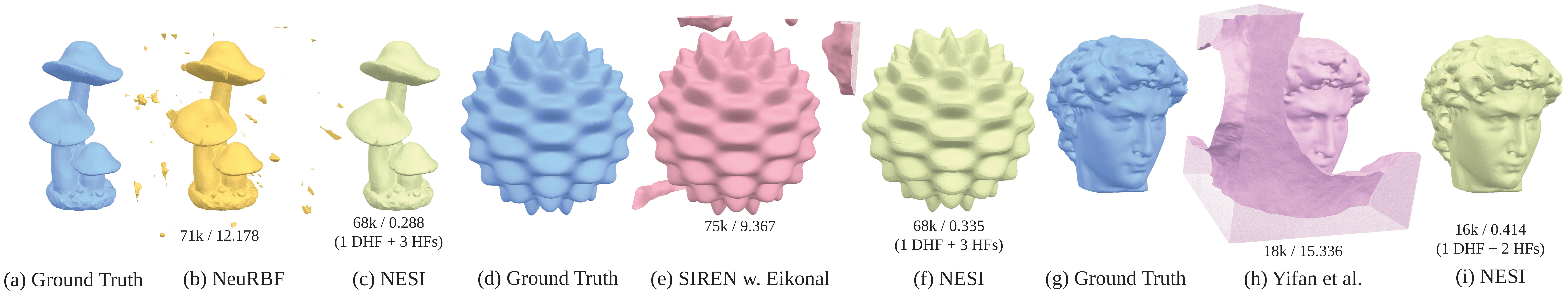}
	\caption{Comparisons of NESI outputs to these of NeuRBF \cite{chen2023neurbf}, SIREN with Eikonal constraints \cite{siren}, and the method of Yifan~\etal~\shortcite{idf}. While NESI results are consistently stable, these prior methods generate spurious extra surfaces on a significant percentage of inputs.}
	\label{fig:floaters}
\end{figure*}

We compare our method against multiple representative state of the art neural implicit methods: SIREN~\cite{siren}, both with and without Eikonal constraints; NeuRBF~\cite{chen2023neurbf}; implicit displacement fields (IDF)~\cite{idf}; NGLOD~\cite{nglod}; VQAD~\cite{vqad}; and the key spheres method of Li et al. \shortcite{spheres}. For all the methods, except Li et al. \shortcite{spheres} we ran the code provided by the authors on the complete dataset detailed above with parameter counts comparable to ours. We were unable to run the code of \cite{spheres}; we therefore qualitatively and quantitatively compare our results to the output models provided by the authors for the Thingi32 dataset. 

Fig.~\ref{fig:floaters} shows representative comparisons of NESI against SIREN with Eikonal constraints \cite{siren}, NeuRBF \cite{chen2023neurbf}, and IDF~\cite{idf}. 
All three methods generate implicit SDFs as their output; \cite{chen2023neurbf,idf} target much larger models than us (700K and 800K parameters respectively) but can be modified to use lower parameter counts. 
On the examples shown (Fig~\ref{fig:floaters}), as well as on many additional inputs, these methods produce unstable outputs with additional spurious zero level-set surfaces. In our experiments, such spurious surfaces appeared in over 50\% of the experiments for NeuRBF and over 30\% for SIREN. For IDF \cite{idf}, the failure rate was highly parameter-count dependent: e.g. for 14K, 18K and 100K parameter counts, the method introduced spurious surfaces on over 30\% of the inputs, but performed well for other counts (e.g. 25K). By using explicit rather than implicit representations, NESI robustly and consistently generates outlier-free approximations. The average and maximal errors (chamfer distance between an output and input shapes) across all input and parameter count combination tested for these methods are 16.5 and 61 for NeuRBF, 6.1 and 81 for SIREN w/eikonal, and 4.47 and 62.5 for \cite{idf}. Our respective average and maximum errors are 0.36 and 1.49, demonstrating NESI's consistency/robustness.

\begin{figure*}
	\includegraphics[width=0.85\linewidth]{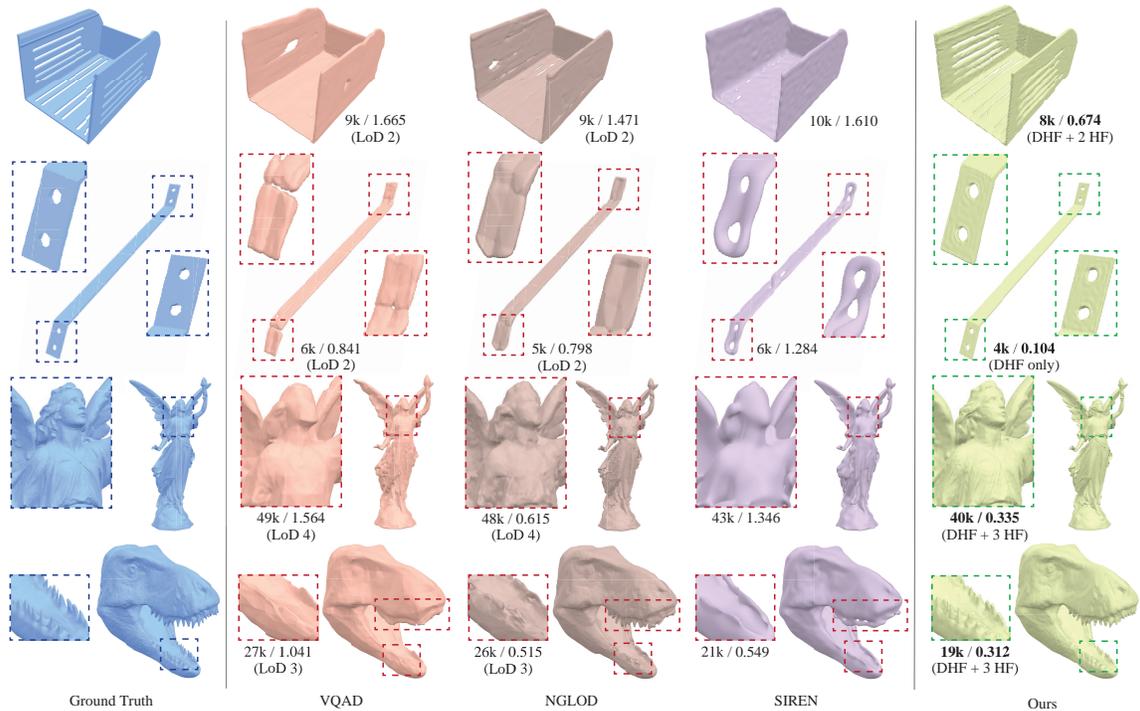}
	\caption{Representative comparisons of NESI outputs to these of SIREN \cite{siren} (no Eikonal constraints), VQAD \cite{vqad}, and NGLOD \cite{nglod}. For similar parameter counts NESI consistently better captures fine details of the input shapes.}
	\label{fig:qual_all}
\end{figure*}

\begin{figure*}
	\includegraphics[width=0.9\linewidth]{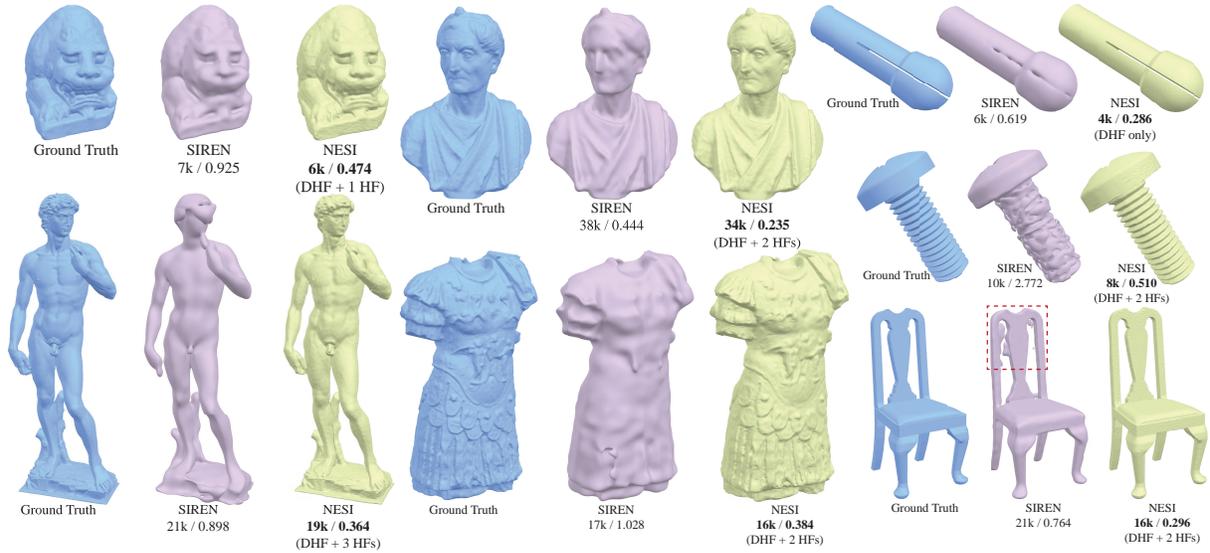}
	\caption{Additional comparisons with SIREN (no Eikonal constraints). Even with smaller parameter counts NESI consistently captures more geometric details.}
	\label{fig:siren_qual}
\end{figure*}

\begin{figure*}
	\includegraphics[width=0.9\linewidth]{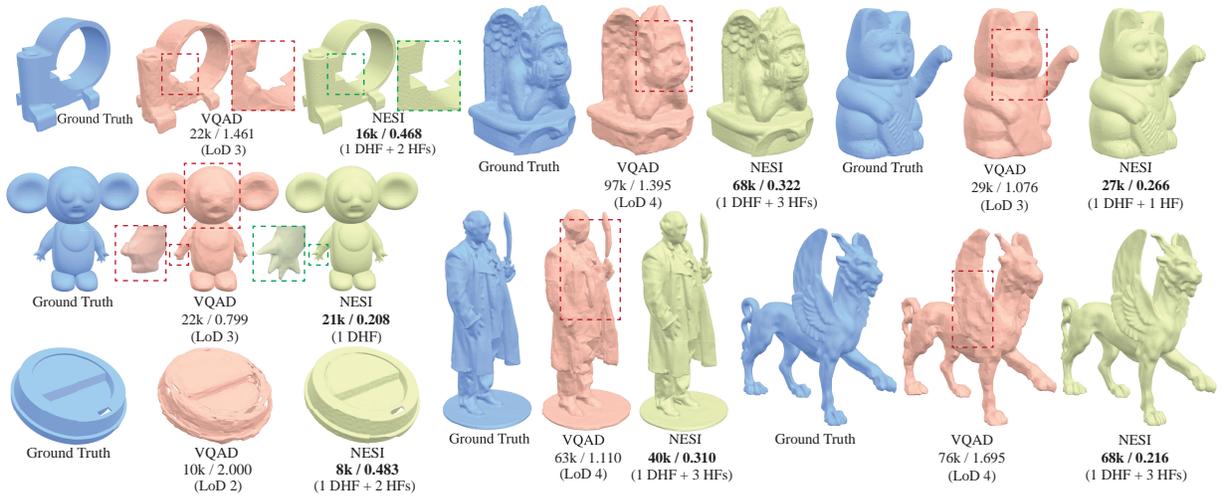}
	\caption{Additional comparisons with VQAD. Even with smaller parameter counts NESI has higher accuracy and captures more details.}
	\label{fig:vqad_qual}
\end{figure*}

\begin{figure*}
	\includegraphics[width=0.9\linewidth]{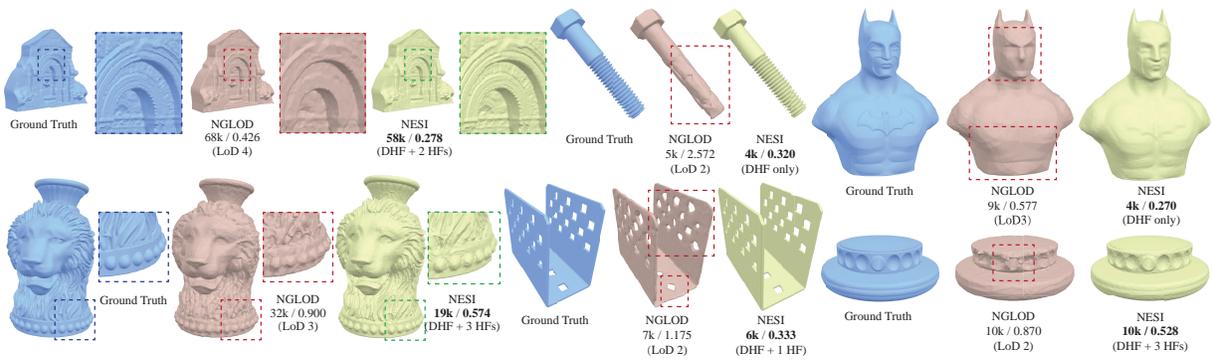}
	\caption{Additional comparisons with NGLOD. Even with smaller parameter counts NESI consistently captures more geometric details.}
	\label{fig:nglod_qual}
\end{figure*}

\begin{figure*}
\includegraphics[width=0.9\linewidth]{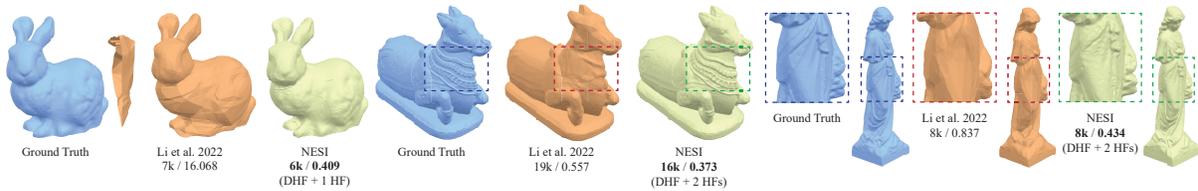}
	\caption{Comparisons with \cite{spheres}. Even with smaller parameter counts NESI consistently captures more geometric details; improvement is most noticeable at lower parameter counts. As the bunny (left) illustrates the method of \cite{spheres} exhibits occasional instabilities while NESI remains robust.}
	\label{fig:spheres_qual}
\end{figure*}

\begin{table}
\begin{center}
	\tiny
	\begin{tabular}{@{}l c c c c c c c@{}}
	Dataset & \#Params &  VQAD & Li et all & NGLOD & \begin{tabular}{@{}c@{}}SIREN\\ (wo. Eikonal)\end{tabular} & NGF &  {\bf NESI} \\[-1pt]
\hline %
ABC      & <10k 	& 2.11 & 	  	& 1.88 & 1.52  & 0.78  & \textbf{0.39} \\[-1pt]
Thingi32 & <10k 	& 2.97 & 2.93	& 3.09 & 1.54  & 0.64  & \textbf{0.54} \\[-1pt]
Others   & <10k		& 2.53 & 	  	& 2.10 & 1.51  & 0.64  & \textbf{0.49} \\[-1pt]
\textbf{All  }    & \textbf{<10k	}	& 2.52 & 	  	& 2.34 & 1.52  & 0.70  & \textbf{0.46} \\[-1pt]
\hline %
ABC      & 10k-20k 	& 1.10 &		& 0.60 & 0.81  & 0.63	 & \textbf{0.41} \\[-1pt]
Thingi32 & 10k-20k 	& 1.55 & 0.67	& 0.78 & 0.82  & 0.55  & \textbf{0.37} \\[-1pt]
Others   & 10k-20k 	& 1.69 & 	  	& 0.85 & 0.83  & 0.61  & \textbf{0.41} \\[-1pt]
\textbf{All }     & \textbf{10k-20k }	& 1.38 & 	  	& 0.72 & 0.82  & 0.60  & \textbf{0.39} \\[-1pt]
\hline %
ABC      & 20k-40k 	& 1.05 & 	  	& 0.51 & 0.59  & 0.46  & \textbf{0.29} \\[-1pt]
Thingi32 & 20k-40k 	& 1.38 & 0.52	& 0.60 & 0.61  & 0.36  & \textbf{0.30} \\[-1pt]
Others   & 20k-40k 	& 1.21 & 	  	& 0.60 & 0.60  & 0.44  & \textbf{0.28} \\[-1pt]
\textbf{All }     & \textbf{20k-40k} 	& 1.20 & 	  	& 0.56 & 0.60  & 0.42  & \textbf{0.29} \\[-1pt]
\hline                                                       
ABC      & >40k 	& 0.73 & 	  	& 0.47 & 0.45  & 0.41  & \textbf{0.35} \\[-1pt]
Thingi32 & >40k 	& 1.01 & 0.50	& 0.49 & 0.48  & 0.33  & \textbf{0.28} \\[-1pt]
Others   & >40k 	& 1.01 & 	  	& 0.48 & 0.47  & 0.36  & \textbf{0.28} \\[-1pt]
\textbf{All}      & \textbf{>40k} 	& 0.90 & 	  	& 0.48 & 0.46  & 0.37  & \textbf{0.30} \\[-1pt]
\hline
\textbf{Overall}  &			& 1.75 &		& 1.40 & 0.93  & 0.51 &  \textbf{0.36}
\end{tabular}
\end{center}
\caption{Quantitative comparison of NESI (right) against prior art \cite{siren,nglod,vqad,ngf,spheres} across different parameter counts and datasets (Chamfer-$L_1$ distance between input and predicted surfaces). NESI consistently outperforms all baselines, with improvement most pronounced for low parameter counts.  
}
\label{tab:allres}
\end{table}

\begin{table}

\begin{center}

	\tiny
    \begin{tabular}{l c c c c c c}
    
	Dataset & \#Params & NESI vs  VQAD  & \begin{tabular}{@{}c@{}}NESI vs\\Li et al.\end{tabular} & NESI vs NGLOD & \begin{tabular}{@{}c@{}}NESI vs\\SIREN (wo.Eikonal)\end{tabular} & NESI vs NGF \\[-1pt]
	\hline
	ABC      & <10k				&  100\% 		& 				&100\% 			&  97\% 		&  83\% 			\\[-1pt]
Thingi32 & <10k 			&  100\% 		& 100\% 		&100\% 			& 100\% 		&  95\% 			\\[-1pt]
Others   & <10k 			&  100\% 		& 				&100\% 			& 100\% 		& 100\% 			\\[-1pt]
\textbf{All} & \textbf{<10k} 	& \textbf{100\%}&				&\textbf{100\%}	& \textbf{99\%} & \textbf{91\%} 	\\[-1pt]
\hline                                                          
ABC      & 10k-20k 			&   96\% 		& 				& 95\% 			& 93\% 			&  90\% 			\\[-1pt]
Thingi32 & 10k-20k 			&  100\% 		& 88\%			&100\% 			& 97\% 			&  91\% 			\\[-1pt]
Others   & 10k-20k 			&  100\% 		& 				& 93\% 			& 93\% 			&  85\% 			\\[-1pt]
\textbf{All} & \textbf{10k-20k} & \textbf{98\%} &				&\textbf{96\%} 	& \textbf{94\%} & \textbf{89\%}		\\[-1pt]
\hline                                                      
ABC      & 20k-40k 			&  100\% 		& 				& 97\% 			& 83\% 			&  88\% 			\\[-1pt]
Thingi32 & 20k-40k 			&  100\% 		& 97\% 			& 95\% 			& 97\% 			&  81\% 			\\[-1pt]
Others   & 20k-40k 			&  100\% 		& 				&100\% 			& 93\% 			&  82\% 			\\[-1pt]
\textbf{All} & \textbf{20k-40k} & \textbf{100\%}&				&\textbf{97\%} 	& \textbf{90\%} & \textbf{84\%}		\\[-1pt]
\hline                                                      
ABC      & >40k 			&   100\% 		& 				&83\% 			&  80\% 		&  78\% 			\\[-1pt]
Thingi32 & >40k 			&  100\% 		& 100\% 		&100\% 			& 100\% 		&  88\% 			\\[-1pt]
Others   & >40k 			&  100\% 		& 				&100\% 			&  93\% 		&  82\% 			\\[-1pt]
\textbf{All} & \textbf{40k}		& \textbf{100\%}& 				&\textbf{94\%} 	& \textbf{90\%} & \textbf{82\%} 	\\[-1pt]
\hline                                                      
\textbf{Overall} & 				& \textbf{100\%}& \textbf{96\%} &\textbf{96\%} 	& \textbf{93\%} & \textbf{86\%} 	\\[-1pt]
	\end{tabular}

\end{center}
\caption{Percentage of shapes that NESI outperforms each baseline \cite{nglod,vqad,siren,ngf,spheres} on across different datasets and parameter counts. Each individual comparison evaluates a NESI output against an alternative method's output produced using same or higher parameter count.}
\label{tab:allrespercent}
\end{table}

We compare NESI both qualitatively (Fig~\ref{fig:qual_all}-\ref{fig:spheres_qual} and quantitatively (Tab~\ref{tab:allres},~\ref{tab:allrespercent}) against four recent methods that use neural implicits: SIREN~\cite{siren} {\em without} Eikonal constraints, VQAD \cite{vqad}, NGLOD \cite{vqad}, and \cite{spheres}. 
SIREN provides an important baseline, since we use the SIREN network (i.e., an MLP with sinusoidal activation functions) as the backbone architecture for neural encoding. The difference is that we use the SIREN network for learning the simple, explicit height functions for different pieces of our NESI representation, while the original SIREN method uses the same network for computing a single zero-level set in $\mathbb{R}^3$ to approximate the entire surface of  an input shape, which can potentially be very complex. VQAD, NGLOD, and \cite{spheres} specifically target low parameter count compression as an application, and thus serve as a natural baseline to our method. These methods were shown to outperform earlier alternatives such as \cite{instantngp} and \cite{davies2021effectiveness} in terms of quality/compactness tradeoff.
Both visual inspection and quantitative comparisons confirm that NESI significantly outperforms all four alternatives,  both on average (Tab.~\ref{tab:allres}) and in head to head comparisons (Tab~\ref{tab:allrespercent}). Our error is 3.8 times smaller than that of NGLOD, 4.9 times smaller than that of VQAD, and 2.6 times smaller than that of the SIREN baseline. It is 1.9 times smaller than that of \cite{spheres} (on Thingi32). 

For the head-to-head comparisons, we encode each input using NESI and an alternative method where NESI encoding uses same or smaller parameter count than the alternative (for each output of an alternative method, we locate our result with the closest parameter count smaller than the alternative). As Tab. ~\ref{tab:allrespercent} shows, NESI outperforms NGLOD on 96\% of inputs tested, VQAD on 100\% of inputs,  SIREN on 93\%, and \cite{spheres} on 96\%. Lastly and critically, NESI is notably more stable than these alternatives: our maximal deviation from ground truth across all inputs and parameter counts is 1.49, versus 6.13 for SIREN, 16 for \cite{spheres}, 29.43 for VQAD, and 35.69 for NGLOD.

\begin{figure*}
\includegraphics[width=0.9\linewidth]{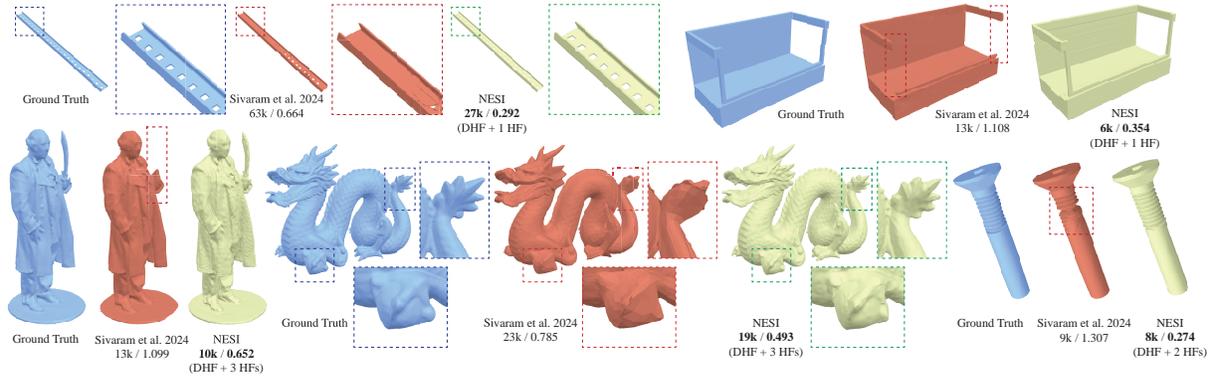}
	\caption{Comparisons with \cite{ngf}. While at lower parameter counts, the alternative method frequently fails to capture prominent topological and geometric features, NESI fatefully approximates these.  } 
	\label{fig:ngf}
\end{figure*}

Finally, we compare our method to the state-of-the-art Neural Geometry Fields (NGF) method \cite{ngf}, that encodes shapes as a combination of a QSLIM simplified mesh and a learned displacement (Fig~\ref{fig:ngf}). Their method fails to produce an output on 21 of the input model/parameter combinations we tested, with most failures happening on models with non-trivial topology ({\em fertility, happy buddha, david, xyzdragon}) at lower parameter counts (QSLIM mesh of 100 to 250 faces). This seems to be due to failures during their triangle pairing step. Visual inspection and quantitative analysis of  their successful outputs confirm that NESI outperforms this alternative, both on average (Tab.~\ref{tab:allres}) and in head-to-head comparisons (Tab~\ref{tab:allrespercent}).
The improvement is most pronounced at lower parameter counts: when using under 10K parameters our average distance is 0.46 versus 0.7 for NGF. A qualitative advantage of NESI over NGFs is its support for straightforward in-out queries and 2D parameterization - neither of which are directly supported by the NGF representation, as it is based on displacement maps.

Please see supplemental material and appendix for additional visual comparisons against the above methods; these comparisons demonstrate that NESI encodings are consistently more accurate and more detailed compared to the alternatives across different parameter counts, with improvement being most noticeable at lower parameter counts.

\subsection{Ablation Studies}
\label{sec:exp_ablation}

We ablate several key algorithmic choices made in ESI and NESI computation.

\begin{figure}
\includegraphics[width=0.9\linewidth]{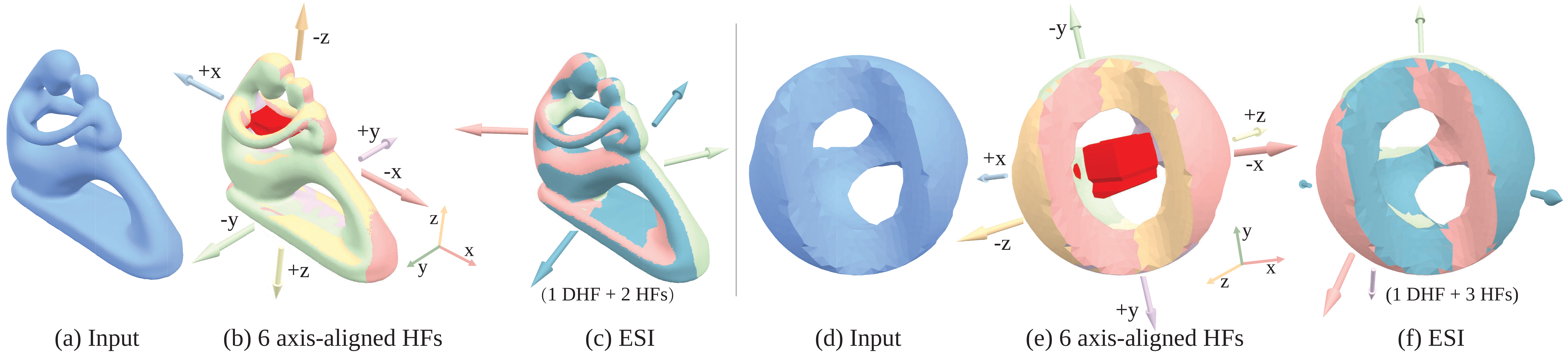}
\caption{Using coordinate system axes as VE axes (6 HFs) (b,e) produces catastrophically poor approximations that dramatically  deviate  (deviating part highlighted in red) from the input (a,d). Our ESIs use notably fewer bounding height fields to accurately approximate these inputs (c,f).}
\label{fig:more_fixed}
\end{figure}

\paragraph{Fixed vs Computed VE axes}
We ablate our strategy for computing DHF/HF axes by comparing it to using a fixed, input independent, set of axes, mimicking depth fusion \cite{shade1998layered,Richter2018Matryoshka} (Fig~\ref{fig:fixed},~\ref{fig:more_fixed}).
As shown in Fig.~\ref{fig:fixed}a-c, having axes in the horizontal plane only \cite{shade1998layered} is clearly insufficient to achieve reasonable approximation even for simple shapes. Similarly, examples in Fig~\ref{fig:fixed},~\ref{fig:more_fixed} demonstrate that using 6 coordinate system axes can lead to catastrophically poor approximation quality. Our method has no such catastrophic failures, despite using notably fewer HFs (maximum 1 DHF + 3HF, average 1 DHF + 1.5 HFs).

\begin{figure}
	\includegraphics[width=\linewidth]{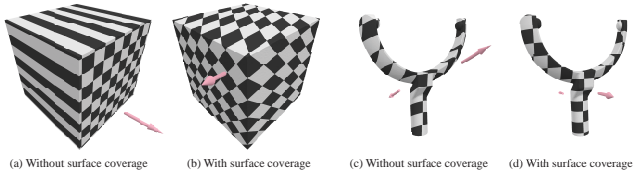}
	\caption{Absent an explicit surface coverage term ESI axis selection may produce volumetric explicits with large surface areas (near)-orthogonal to all VE axes (a, c). While the resulting ESIs are still suitable for in-out queries, parameterizing these areas via projection leads to extreme parametric distortion. (b, d) Our loss function (Eq~\ref{eq:esi}) which accounts for coverage, results in outputs that can be parameterized with much lower distortion.}
	\label{fig:ortho}
\end{figure}

\paragraph{ESI Axis Selection.}
Our ESI optimization explicitly accounts for surface coverage  (Eq. ~\ref{eq:esi}) and promotes outputs that can be projectively parameterized without excessive distortion or degeneracies. Fig.~\ref{fig:ortho} demonstrates the impact of eliminating this penalty: while the resulting DHF approximation (Fig.~\ref{fig:ortho}b) precisely captures the input geometry, its projection along the axis is not bijective. Our output (Fig.~\ref{fig:ortho}c) is parameterized bijectively, facilitating texturing. Fig.~\ref{fig:chair_spiral} validates another key choice of our method: the use of symmetric rather than one sided distance between the input and the ESI approximation. While distance from ESI-to-input is much more computationally expensive to compute than the inverse, using it as part of the optimized loss function is critical for obtaining suitable approximations. 

\begin{figure}
	\includegraphics[width=0.7\linewidth]{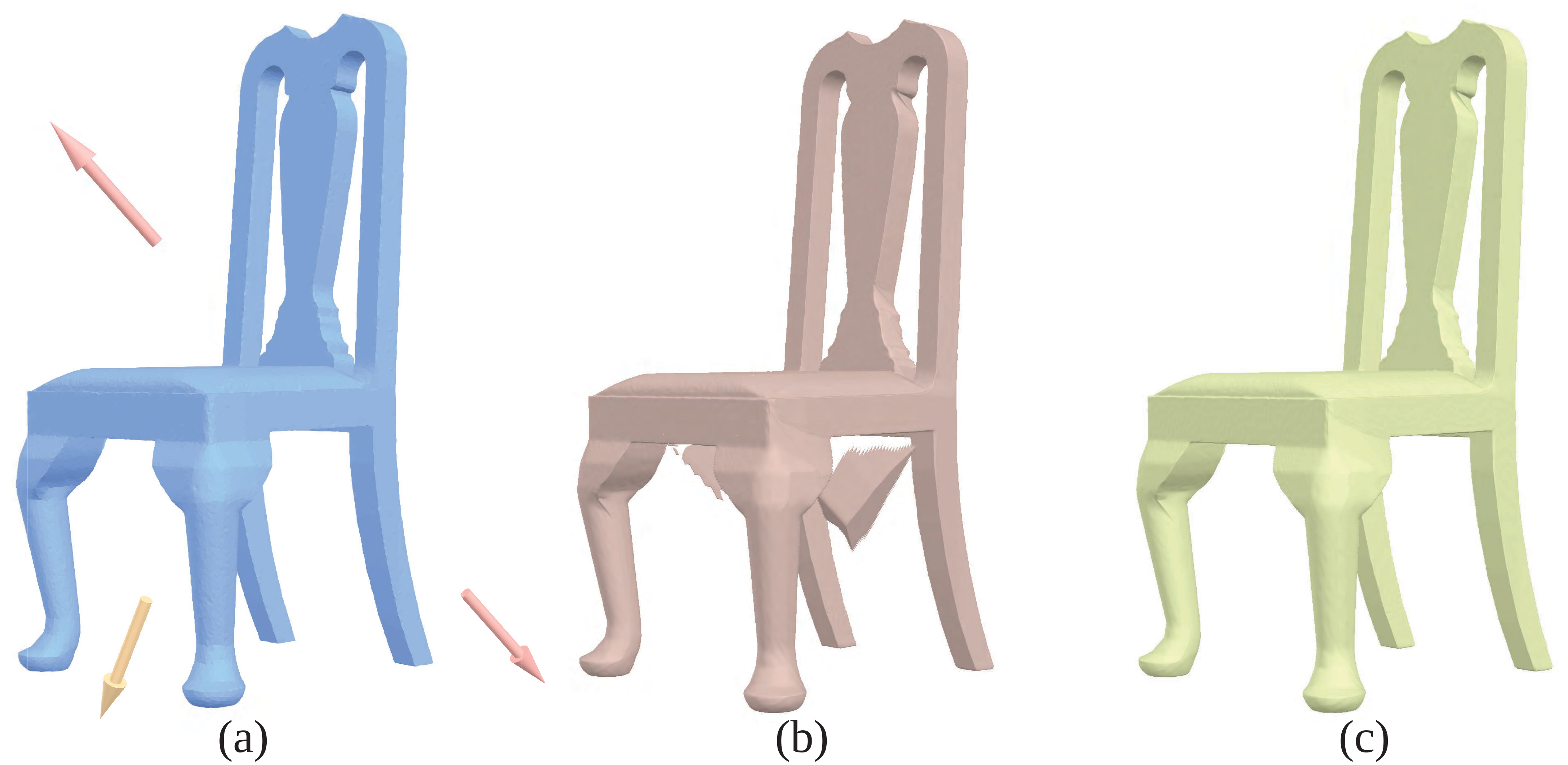}
\caption{Given the input (a, arrows illustrate DHF and HF axis directions), using the method in Sec. 5 to define $\tilde \Omega_k$ for $k=1,2$ produces the desired outcome (c); removing the ray visibility condition for HFs (\#3) results in an output that does not accurately capture the void between the chair legs (b).}
\label{fig:choices}
\end{figure}

\paragraph{NESI Computation.}
Figure~\ref{fig:choices} demonstrates the impact of our set of tests used to define $\tilde \Omega$ (Sec~\ref{sec:nesi}). Using a more restrictive definition of $\tilde \Omega$ (one that omits the ray visibility test) can produce outputs that may not well approximate interior voids (Fig~\ref{fig:choices}b). This criterion can be thought of as mirroring the need to measure the distance from ESI to input in the axis computation above.  

\paragraph{Localized HFs.} We reduce NESI memory footprint and avoiding learning duplicate surface geometry by learning HF surface geometry only in areas where it is not already covered by the DHF/other HFs (Fig~\ref{fig:nesi}de, Sec~\ref{sec:learning}). We ablate the importance of this step by comparing the accuracy of our results to that of results generated using same parameter counts, but without such localization (using $\tilde \Omega = \Omega$) on inputs that require at least one HF (localization is not performed when only a single DHF is used). Using our approach improves approximation quality by 15\% across all parameter counts (0.39 NESI vs 0.45 ablation), with the improvement being more pronounced for small parameter counts - at parameter counts under 10K, our choice leads to a 25\% improvement over the ablation alternative (0.55 NESI, vs. 0.69 ablation). 

\paragraph{Normal Preservation.} We measured the difference in approximation quality between results obtained using our loss functions $\loss^{D}, \loss^{H}(k)$ (Eqs~\ref{eq:loss_dhf}, \ref{eq:loss_hf}) and ones which do not include the normal terms $\mathcal{L}_{Normal}^{H}(k),\mathcal{L}_{Normal}^{D}$
(Eqs~\ref{eq:normal_dhf}, \ref{eq:normal_hf}). While in theory one can expect Chamfer distance to improve when normals are not optimized for, we found that including the normal terms marginally reduces approximation error on average (CD 0.364 with normal preservation and 0.367 without; adding the term leads to better approximation on 288 out of 400 inputs).  Visual assessment confirms that our results look better than ones without the normal preservation term. 

\paragraph{Image Compression.} A naive alternative to our learned neural representation of ESI, is to convert the HFs and the two DHF height-fields into depth images and store those in compressed form.  To ablate this alternative, we stored the ‘batman’  (Fig. ~\ref{fig:nglod_qual}) DHF and HFs as 500$\times$500 depth-images (JPEG; 300K filesize). Decompressing and combining these HFs produces an ESI model with chamfer distance of 0.36 to input; a NESI output with 1/10 the filesize (24K, 6K parameters) has chamfer distance of 0.27. This experiment demonstrates that this alternative provides a much worse size/accuracy tradeoff. Moreover, while NESI can be processed {\em as-is}, compressed images need to be decompressed before processing with decompression significantly increasing their memory footprint. Our ability to support as-is processing distinguishes NESI from all traditional image and geometry compression methods which require decompression before processing.

\paragraph{File Size.} The file size of a NESI model is determined by two factors: the number of explicits used and the number of parameters allocated to each.
Our file sizes range from 24K for a single 6K parameter DHF, to 273K for a 38K parameter DHF and three 10K parameter HFs.  

\paragraph{Runtimes.} All models in our experiments were trained on an NVIDIA Tesla V100 16GB for 10,000 iterations, which takes between 5 and 30 minutes depending on the number of explicits used and network size. Determining the optimal number of explicits and their axes takes on average 3 minutes, and up to 12 minutes for our worst performing model ({\em hundepaar} from the DHFSlicer dataset). 

\subsection{Discussion and Limitations.}
\label{sec:discuss}

\paragraph{Subtractive Formulation of NESI.} It is instructive to take an alternative view of the Boolean definition of the NESI representation based around the restricted domains $\tilde \Omega_k$ for each HF. Recall that the NESI representation is defined as the approximating volume

\begin{equation} 
\tilde S = DHF({\bf d_0}) \cap (\cap_{k=1}^m HF_k ({\bf d}_k).
\end{equation}
Clearly, it can be re-written as 
\begin{equation} 
{\tilde S}  = DHF({\bf d_0}) \setminus (\cup_{k=1}^m \overline{HF}_k ({\bf d}_k),
\label{Eqn_HF}
\end{equation}
where $\overline{HF}_k ({\bf d}_k) = \{ (x,y,z) | z \geq f_k (x,y) \wedge (x,y) \in \Omega_k\}$.

We now observe that if we adopt the Boolean definition of ${\tilde S}$ in Eqn. \ref{Eqn_HF}, each $\overline{HF}_k$ can be viewed as a subtractive volumetric primitive. It then follows that this Boolean expression of ${\tilde S}$ in Eqn. \ref{Eqn_HF} holds true when we replace the domain $\Omega_k$ of $\overline{HF}_k$ by the restricted domain ${\tilde \Omega}_k$, by our construction. 

\paragraph{NESI Scope/Limitations.} NESI representation is targeted at consumer facing graphics applications, such as video games, online shopping, AR/VR, or remote communication. As such we aim to create representations of typical everyday shapes that have very low memory footprint, and that can be effectively rendered on consumer devices.   

\begin{figure}
	\includegraphics[width=\linewidth]{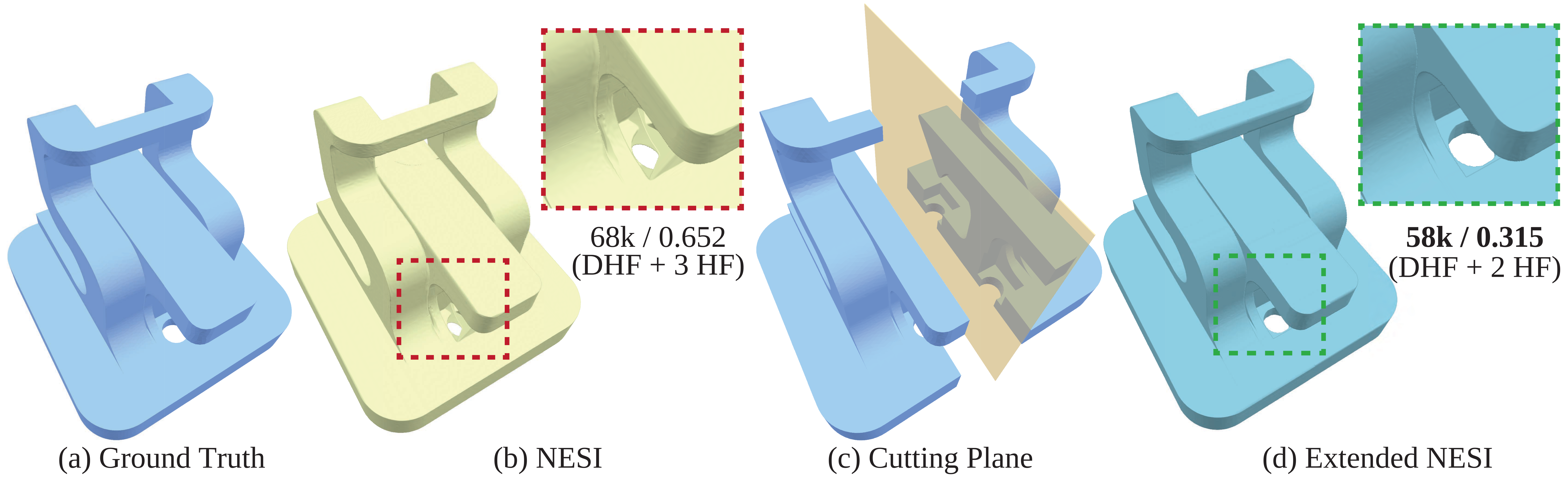}
	\includegraphics[width=\linewidth]{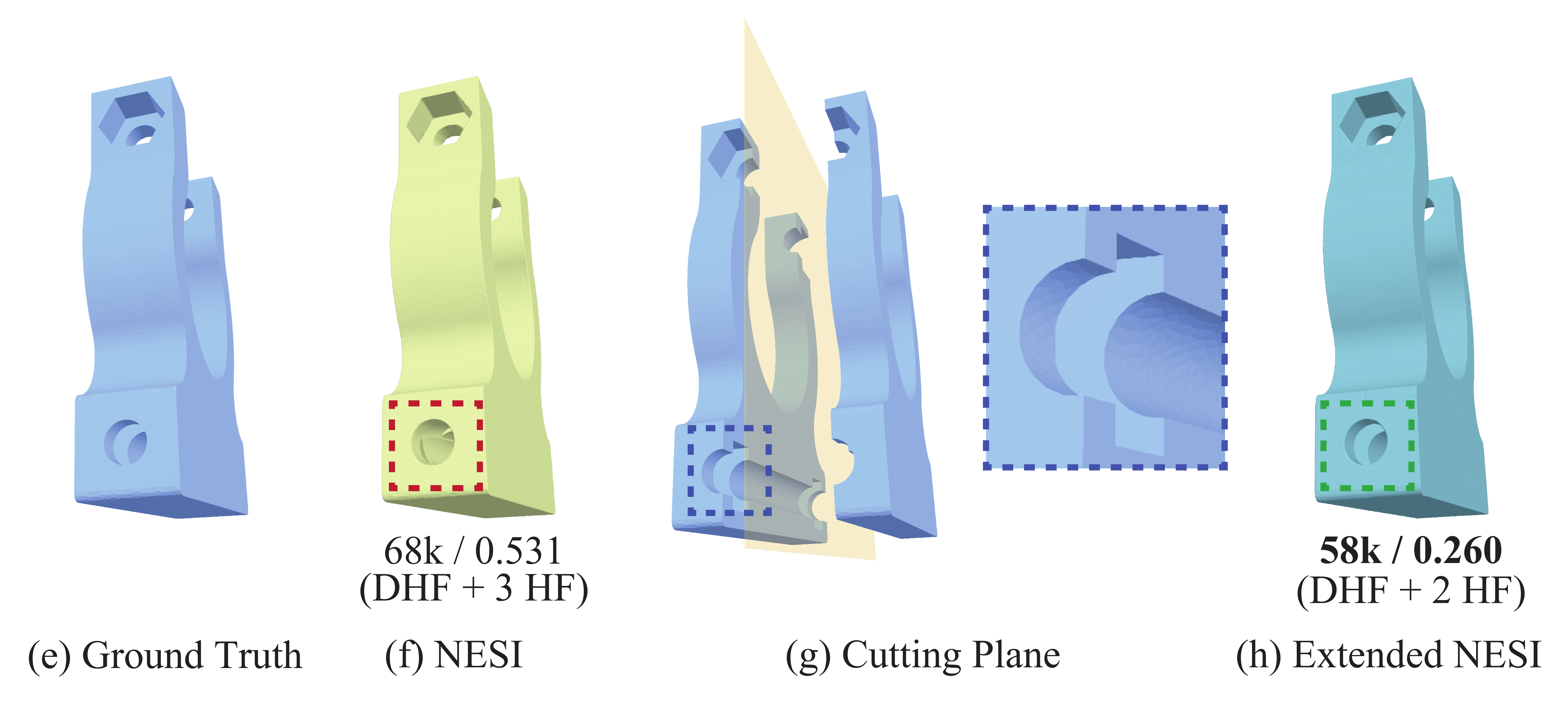}
\caption{Shape with severe self-occlusions (a,e) may require a very large number of volumetric explicits to obtain a numerically accurate approximation (interior voids entirely invisible from outside simply cannot be captured). However, for practical purposes, using a DHF hull and 3 HFs already provides a good approximation of the shape for rendering from any external viewpoints (b,f). Using our extended method (c,d,g,h) improves numerical accuracy for such inputs. Notably while numerical accuracy improves - visually both the default (b,f) and the extended method outputs (d,h) look very similar to the inputs (a,e).}
\label{fig:limitations}
	\label{fig:extension}
	\label{fig:extensions}
\end{figure}
 
Consequently, like other implicit representations \cite{nglod,vqad,idf}, our method is, by construction, designed for closed objects that can be well represented in implicit form. This constraint makes it well suited for representing most human-made and organic content, but less suited for objects such as leafy trees or clothing.

NESI is based on capturing surface geometry visible from outside the processed shapes. This assumption is consistent with the typical rendering setup in the applications we target. 
Given shapes with surface regions that are heavily or entirely occluded (Fig~\ref{fig:extension}ad), NESI approximations will still produce visually high quality results when rendered from outside (Fig~\ref{fig:extension}be) but will exhibit high numerical error. The approximation quality can be further improved by using our extended method  (Sec.~\ref{sec:extensions}, Fig.~\ref{fig:limitations}dh).
Since such surface regions almost never need to be rendered, this limitation is rarely relevant for practical settings for viewing purposes. We note that all numerical results reported in the paper as well as all other visuals do {\em not} include this extension.

\section{Conclusions}
\label{sec:conclude}

We presented NESI, a novel compact neural representation for 3D shapes.  Our representation combines the processing advantages of implicit and parametric representations, making it exceptionally suitable for a wide range of geometry processing applications. 
Our experiments %
convincingly demonstrate that NESI approximates diverse, complex 3D shapes much more accurately than state-of the art alternatives, when using the same parameter count, or memory footprint. This improvement is  most pronounced at lower parameter counts, where our average error is 35\% smaller than that of closest alternative (0.46 NESI vs 0.7 \cite{ngf}). Across the entire set of inputs and parameter counts NESI outperforms this closest alternative 86\% of the time. 
This improvement is made possible by our reduction of the 3D approximation problem to a combination of two sub-problems: (1) locating optimal DHF and HF axes such that the intersection of these volumetric explicits tightly approximates the input; and (2)  compactly representing each DHF and HF as a 2D neural function such that the intersection of these functions well approximates the input.

\paragraph{Future Work.}
One interesting future NESI direction is to explore the use of explicit surfaces defined over non-Euclidean spaces, where the mapping between the domains and surfaces is not necessarily orthographic but is perhaps perspective or fish-eye based. Using such surfaces has the potential to further reduce the number of explicits necessary to accurately represent complex shapes. Another is to explore additional geometry processing applications that can benefit from our ESI and NESI representations. Given that ESIs are guaranteed to always contain the input shape, one potential, immediate, application is collision detection.

\bibliographystyle{ACM-Reference-Format}
\bibliography{ref}
\appendix
\section{Method Implementation Details.}

\paragraph{Network Architecture.} For our experiments, we represent DHFs as MLPs with 5 hidden layers; all activation functions are sinusoidal except for the last layer, which has no activation functions. We allow the width of each layer to change depending on the desired tradeoff between compactness and accuracy; in our experiments, we used four different settings for layer width (32, 48, 72, 96) depending on the desired parameter count. We encode HFs as two MLPs; one MLP for the height function, and one smaller MLP for the mask. The HF uses 5 hidden layers, with four different settings for layer widths (16, 24, 36, 48) depending on the target parameter count. We note that, for a given compression setting, the DHF layer width is exactly double the HF layer width. The mask MLP returns a binary function $\{0,1\}$ which indicates whether the input point is inside or outside the domain $\Omega$; internally, the MLP returns a value clamped by a sigmoid function to lie between $[0,1]$, and we say that a point is inside the mask if the MLP returns 0.5 or greater. The mask MLP has 3 hidden layers, with 16 as its width for all experiments. 

All parameters in our experiments are represented as 32-bit floats. All modules are trained with an ADAM optimizer, using a learning rate that starts at $10^{-3}$ and decays to 0 using a cosine annealing scheduler. In each iteration we randomly select 500k points from the set of sampled points, with 250k in the surface region $\Omega$ and the other half outside it. For HF networks, we further split the surface training samples between 125k in $\tilde\Omega$ and 125k in $\Omega\setminus\tilde\Omega$. 

\paragraph{ESI Computation Parameters.} We set $\varepsilon_{1}=10^{-4}$ and $\varepsilon_{2}=10^{-3}$ empirically to prioritize volumetric approximation over surface coverage. We use 10000 sample points for input and VE discretizations.

\section{Application Implementation Details.}

We provide implementation details of our geometry processing applications below.

\paragraph{Raytracing.} Raytracing requires two operations: computing ray-surface intersections; and computing normals at the points of intersection. While \emph{any} ray-surface intersection algorithm can be used, we use a simple strategy that provides guaranteed error bounds for our evaluation, based on subdivisions. We evenly sample $n$ points on a given ray and check their occupancy values. We then find the first segment whose occupancy changes, starting from the source point.
We refine this segment by evenly subdividing it into $n$ points and repeat this process.
In our quantitative evaluations, we set $n{=}200$ and repeat this process three times such that the final segment is sufficiently small.

\paragraph{Texture Atlassing.}
We can use the parametric representation underpinning NESI by exploiting the mapping from the parameter domains $\Omega_i$ to store input surface signals in parameter space (Fig. \ref{fig:applications}). Given a textured object, we first use the explicit mapping from points on the surface to the explicits that contain that point to save the surface signal in an atlas. At render time, our renderer identifies the explicit(s) that cover the given point and fetches the signal from the atlas. In the example in the paper we use this approach to transfer texture from a mesh to the NESI representation of this mesh, and subsequently render it by raytracing the NESI.

\paragraph{Meshing.} Many downstream applications for neural surface representations require converting them back to a mesh-based format, with the most common approach being an isosurface extraction step (e.g. marching cubes). Because NESIs are defined by explicit surfaces, we can mesh them by exploiting their parametric domain representation, producing superior outputs. 

We start by generating a watertight mesh of the DHF hull by creating a grid of points covering $\Omega_0$, and directly generate a mesh $M_0$ from the DHF hull by assigning DHF hull height values $f_a$ to points on the grid and connecting points with triangles in the obvious manner. We repeat this process for $f_b$, then stitch the two sides together. To ensure that the DHF hull mesh is high quality, we remesh grazing regions (triangles whose normal with respect to the DHF axis is greater than $70^\circ$) that join the two sides of the DHF using an isotropic remesher \cite{botsch2004remeshing}. Then for each $HF_{k}$, we generate a mesh proxy $M_{k}$ that represents the bounded volume between $z=0$ and the height values within the HF's surface domain.  We then create a final mesh by incrementally computing the CSG difference operation of $M_{0}\setminus M_{1}\cdots\setminus M_{k}$. To avoid discretization issues during subtraction in regions where the surface is represented by the DHF and one or more HFs, before performing the subtraction task we find any vertex on $M_{k}$ that is also covered by $M_{0}$ and offset it slightly inwards along its normal. As shown in the paper our meshes closely approximate the input geometry. %

\begin{figure}
	\includegraphics[width=\linewidth]{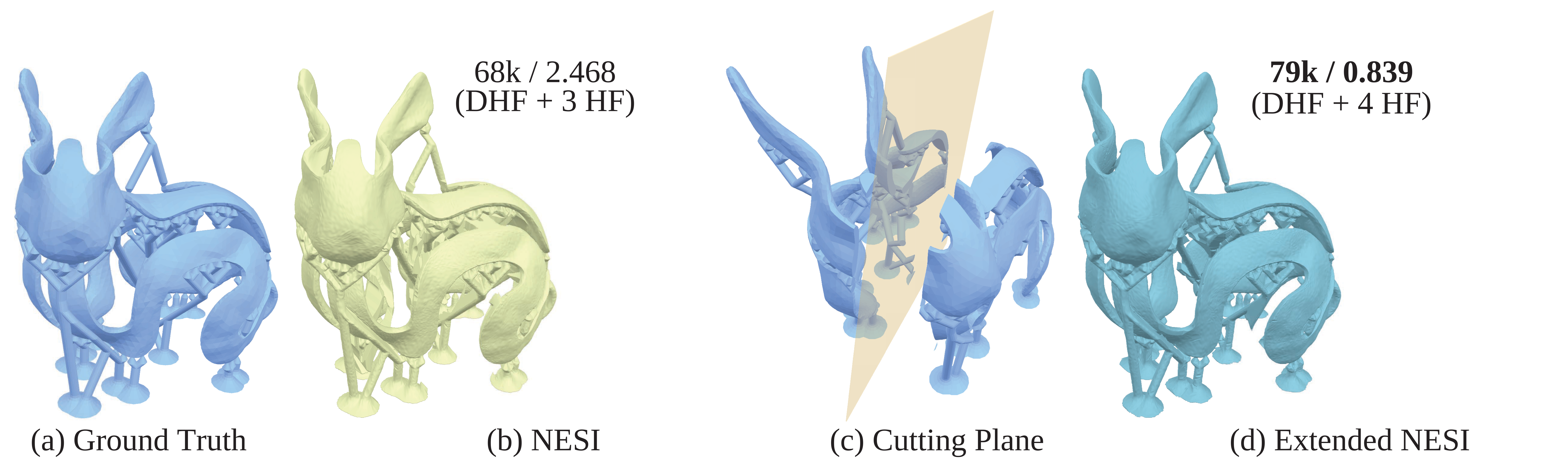}
	\caption{Our extended method combines the DHF of the input with HFs computed on intersections of the input with half-spaces (c) to obtain numerically more accurate results (d) than those produced by the standard method (b). At the same time, from outside the rendered NESI (b) and extended NESI (c) results appear practically identical to one another, and to the input (a).}
	\label{fig:extension2}
\end{figure}

\section{Extension: Handling (Nearly)-Occluded Surfaces}
\label{sec:extensions}
 
By default, NESI targets approximation of surfaces that are visible from some points outside the input shapes. This is consistent with the consumer facing applications we target, where users see content rendered from outside.  As such, NESI is not designed for capturing interior voids or surfaces which are only revealed when a shape is cut open (Fig~\ref{fig:extension2}c). 

In cases where capturing such surfaces is important for application purposes (e.g. renderings from inside such voids) one can use the following BSP-like strategy to capture them.
Specifically, the proposed strategy subdivides the original shape by cutting planes separating pairs of occluded/occluder elements. Starting from the input surface, we test visibility along our candidate DHF directions by sampling multiple points on each face, shooting a ray along the DHF axis direction, and checking for intersections. A face is said to be {\em occluded} if it is not visible from any direction; that is, no matter what candidate direction a ray follows from points on the face, it always hits some other part of the surface geometry. We then find the weighted average of all occluded face vertices, weighted by face area, to find a centroid for a cutting plane; then select a normal direction for the cutting plane from our candidate DHF directions which eliminates the most occlusions. 

We then add two new HFs to the NESI whose directional axis is the normal of the cutting plane but in opposing orientations, and that are situated such that the $x-y$ plane of the HF (i.e. where $z=0$) lies on the cutting plane. Finding sample points for the HF and training then proceeds as usual. When using these HFs for in/out tests, we allow the intersection test to pass if one of the following two conditions are met: either $z \leq f(x,y)$ (the standard criterion); or $z \leq 0$, in which case the point being tested is on the other side of the cutting plane, and so this HF should not be considered when testing this point. Additional HFs are added to each half using the same process as for the standard method. This process can be repeated if multiple cuts are necessary. In practice, we offset the plane locations by a slight offset $\varepsilon$ along the cutting plane directional axis so that the two HF regions overlap. Inserting HF planes in this manner and retaining a common DHF hull, rather than subdividing the NESI explicitly into two, eliminates the risk of false negatives without introducing false positives when performing in-out queries.

Figs. ~\ref{fig:extension2} and ~\ref{fig:limitations} compare the results quantitatively with and without the extension for shapes with occluded surfaces. This extension allows NESIs to capture even highly complicated geometry such as the scaffolded bunny (Fig.~\ref{fig:limitations}) which has many regions that are invisible or barely visible from outside. At the same time, the results with and without the extension look practically identical when rendered from outside. 
We note that all numerical results reported in the paper, as well as all other visuals, do {\em not} include this extension.

\begin{figure*}
	\includegraphics[width=0.9\linewidth]{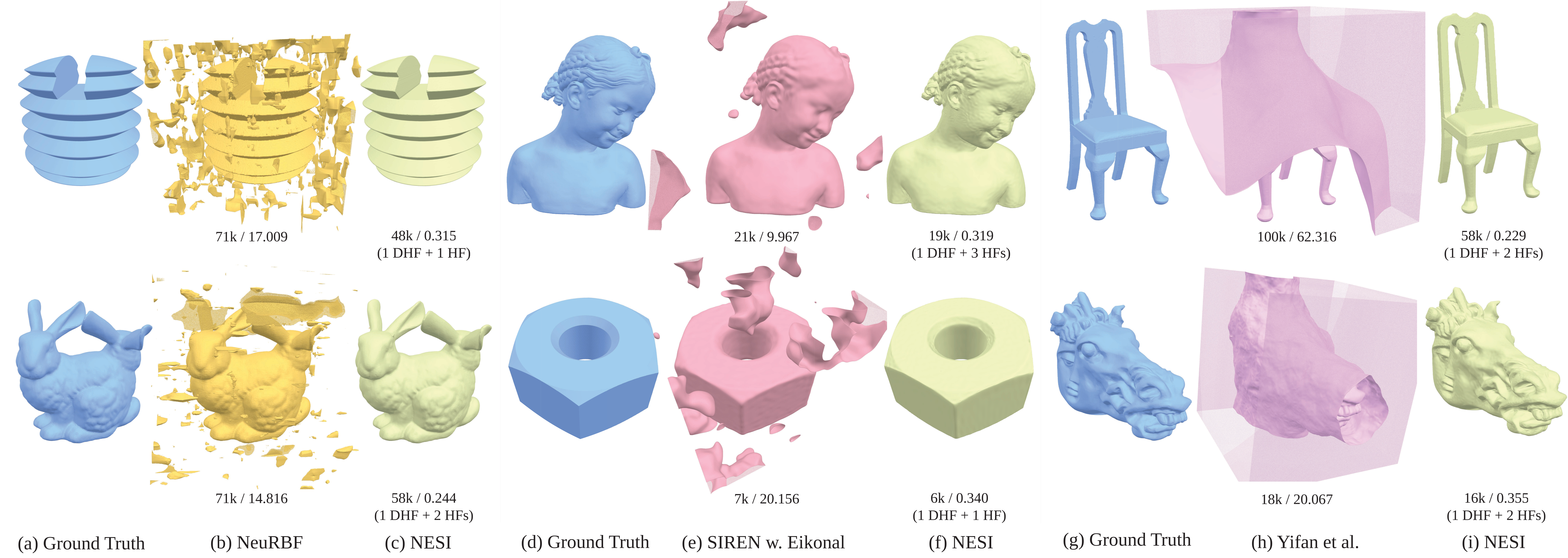}
	\caption{Additional comparisons of NESI outputs to NeuRBF \cite{chen2023neurbf}, SIREN with Eikonal constraints \cite{siren}, and Yifan~\etal~\shortcite{idf}. While NESI results are consistently stable, these prior methods generate spurious extra surfaces on a significant percentage of inputs.}
	\label{fig:floatersmore}
\end{figure*}

\section{Algorithm and Comparison Settings}

\paragraph{Data Sourcing.} We evaluate NESI on 100 diverse shapes taken from four publicly available datasets, chosen to represent a broad range of shapes commonly used in graphics applications and related prior art. We include the Thingi32 subset of Thingi10k~\cite{thingi10k} (32 shapes), used by several prior methods for neural shape representation (e.g. ~\cite{nglod,spheres}), to provide comparison with shapes chosen by the baseline methods. We include 40 semi-randomly selected objects from the ABC~\cite{abcdataset} dataset of CAD models; as this dataset contains many basic (e.g. cubes or cylinders) and repetitive shapes, we curate the subset to keep only relatively complicated shapes and avoid repetition. We also use 25 objects from the DHFSlicer~\cite{yang2020dhfslicer} paper which first introduced the notion of DHF surfaces and experimented with decomposing diverse shapes into DHF blocks. Finally, we include 3 canonical, complex scanned shapes from the Stanford 3D Scanning Repository \cite{StanfordScanRep}: the {\em Buddha}, {\em Thai statue}, and {\em David} models. To evaluate performance of the ESI representation without neural encoding, we augment this data set with 122 additional models from the Thingi10k dataset, chosen at random; and the data set of Myles et al. ~\shortcite{Myles16}, consisting of 98 shapes commonly used in computer graphics.

Two additional models are used to illustrate different aspects of the method: the `spot' cow \cite{crane2013robust} (Fig. 10 in our paper) and the scaffolded bunny (Fig~\ref{fig:extension2}, from the {\em thingiverse} model repository under Creative Commons Attribution). 

\paragraph{Parameter Counts: NESI and Alternatives.} We test our outputs with four network size settings for DHF and HF networks: (1) DHF network with 4390 parameters and respective HF network with 1766 parameters; DHF network with 9654 parameters and respective HF network with 3110 parameters; DHF network with 21390 parameters and respective HF network with 6088 parameters; and DHF network with 37734 parameters and respective HF network with 10214 parameters. Our overall parameter counts depend on the number of HFs used; at our lowest DHF model size, the overall model sizes vary from 4390 (zero HFs) to 9688 (3 HFs), and at the highest model size from 37734 (zero HFs) to 68376 (3 HFs). 

For NGLOD and VQAD we generated models at levels of detail 1 through 4, where their method determines the parameter counts per model based on level of detail. For SIREN we used parameter counts of: 4801, 6841, 9241, 12001, 16241, 21121, 26641, 32801, 39601, 47041, 53041, 61601, and 75641. For \cite{idf} we used parameter counts of 6K, 14K, 18K, 25K, 39K, 56K, and 100K. For NeuRBF we use hash map sizes from 6 to 11. To compare against Sivaram \etal \shortcite{ngf} (NGF), we adjust their ``mesh LOD'' parameter, which in turn adjusts the target number of faces during their QSLIM operation. Numbers used for mesh LOD are taken from their paper, except for 500 and 2000, which we added to have parameter counts that are closer for comparison. We only change NGF's MLP width when comparing on less than 12k parameters.

\paragraph{Comparison Setting}
All renderings of both our and alternative results were generated via our raytracing code and colored using a flat shading scheme. We quantitatively evaluate our method through the standard protocol \cite{nglod} of measuring Chamfer-$\Lone$ distance, measured using 5 million well-sampled points on the input mesh and the output models. We opt for a high sampling rate in order to capture surface-to-surface distances as accurately as possible. Direct comparison across methods with different parameter counts is challenging; we therefore report two sets of numbers in the main paper: the accuracy of each method for models within a fixed parameter range (e.g. under 10K, or 10k to 20K); and pairwise comparisons where, for each output of our method, we algorithmically select the alternative method output with the smallest parameter count that is larger than, or equal to, ours. The latter strategy evaluates the percentage of inputs on which our method outperforms the alternative, while the former quantifies the magnitude of improvement. Both approaches show that that our method significantly outperforms all baselines given the same parameter count.

\section{Additional Visual Comparisons}

Fig. ~\ref{fig:floatersmore} shows additional comparisons of NESI outputs to those of NeuRBF \cite{chen2023neurbf}, SIREN with eikonal constraints \cite{siren}, and Yifan ~\etal~\shortcite{idf}, demonstrating these approaches' lack of robustness.

\end{document}